\documentclass[11pt]{article}

\usepackage[T1]{fontenc}
\usepackage[utf8]{inputenc}
\usepackage{times}
\usepackage[
    font=small,
    labelfont = bf
]{caption}

\usepackage[scale=0.7, marginratio={1:1, 2:3}]{geometry}
\usepackage{setspace}

\usepackage{graphicx}            
\usepackage{float}
\usepackage{amsmath,amssymb} 
\usepackage{mathptmx}          
\usepackage{siunitx}
\usepackage{bm}
\usepackage{mathtools}
\allowdisplaybreaks[3] 

\usepackage{needspace}
\usepackage{etoolbox}
\AtBeginEnvironment{equation}{\Needspace{5\baselineskip}}
\AtBeginEnvironment{equation*}{\Needspace{5\baselineskip}}
\AtBeginEnvironment{align}{\Needspace{6\baselineskip}}
\AtBeginEnvironment{align*}{\Needspace{6\baselineskip}}
\AtBeginEnvironment{gather}{\Needspace{5\baselineskip}}
\AtBeginEnvironment{gather*}{\Needspace{5\baselineskip}}
\AtBeginEnvironment{multline}{\Needspace{6\baselineskip}}
\AtBeginEnvironment{multline*}{\Needspace{6\baselineskip}}
\usepackage{dcolumn}
\usepackage{multirow}
\usepackage[table,x11names]{xcolor}
\usepackage{array}
\newcolumntype{C}[1]{>{\centering\arraybackslash\hspace{0pt}}p{#1}}

\usepackage[caption=false,justification=justified,singlelinecheck=false]{subfig}

\usepackage{xurl}
\usepackage{hyperref}
\usepackage{csquotes} 
\usepackage[backend=biber,style=nature]{biblatex}
\renewbibmacro*{url+urldate}{}
\DeclareFieldFormat{title}{%
  \iffieldundef{url}
    {#1}%
    {\href{\strfield{url}}{#1}}%
}

\usepackage[english]{babel}
\usepackage{algorithm2e}
\usepackage{notoccite}
\usepackage{upgreek}
\usepackage{scalerel}
\usepackage[export]{adjustbox}
\usepackage{textcomp}
\usepackage{placeins}
\usepackage{color}

\renewcommand{\thetable}{\MakeUppercase{\roman{table}}}

\usepackage{authblk}

\title{Social Media Data for Population Mapping: A Bayesian Approach to Address Representativeness and Privacy Challenges}
\date{\today}

\author[1,2]{
    Paolo Andrich
    \thanks{
        Corresponding authors: 
        \texttt{paolo.andrich@cs.ox.ac.uk}, 
        \texttt{shengjie.lai@soton.ac.uk}
    }
}
\author[2]{Shengjie Lai\protect\footnotemark[\value{footnote}]}
\author[3]{Halim Jun}
\author[2]{Qianwen Duan}
\author[2]{Zhifeng Cheng}
\author[1]{Seth R. Flaxman}
\author[2]{Andrew J. Tatem}

\affil[1]{Department of Computer Science, University of Oxford, Oxford OX1 3QG, United Kingdom}
\affil[2]{WorldPop, School of Geography and Environmental Science, University of Southampton, Southampton SO17 1BJ, United Kingdom}
\affil[3]{Global Office of Innovation, United Nations Children's Fund, Stockholm 112 34, Sweden}

\begin{document}

% \bibliographystyle{naturemag}
% \begin{bibunit}[naturemag]
\begin{refsection}

\maketitle

\begin{abstract}
Accurate and timely population data are essential for disaster response and humanitarian planning, but traditional censuses often cannot capture rapid demographic changes. Social media data offer a promising alternative for dynamic population monitoring, but their representativeness remains poorly understood and stringent privacy requirements limit their reliability. Here, we address these limitations in the context of the Philippines by calibrating Facebook user counts with the country's 2020 census figures. First, we find that differential privacy techniques commonly applied to social media-based population datasets disproportionately mask low-population areas. To address this, we propose a Bayesian imputation approach to recover missing values, restoring data coverage for 5.5\% of rural areas. Further, using the imputed social media data and leveraging predictors such as urbanisation level, demographic composition, and socio-economic status, we develop a statistical model for the proportion of Facebook users in each municipality, which links observed Facebook user numbers to the true population levels. Out-of-sample validation demonstrates strong result generalisability, with errors as low as $\approx$18\% and $\approx$24\% for urban and rural Facebook user proportions, respectively. We further demonstrate that accounting for overdispersion and spatial correlations in the data is crucial to obtain accurate estimates and appropriate credible intervals. Crucially, as predictors change over time, the models can be used to regularly update the population predictions, providing a dynamic complement to census-based estimates. These results have direct implications for humanitarian response in disaster-prone regions and offer a general framework for using biased social media signals to generate reliable and timely population data.
\end{abstract}

\section{
\label{sec:intro}
Introduction
}

Each year natural disasters affect the lives of millions around the World and kill, on average, 40,000-50,000 people\cite{Ritchie2022}. Despite this, over recent decades, disaster-related mortality has significantly decreased due to the development of better infrastructures, early warning systems and logistic response mechanisms\cite{Ritchie2022}. These improvements play a crucial role in disaster mitigation by enabling timely evacuation, rescue and assistance operations, and resource allocation. However, the effectiveness of such responses depends on accurate and up-to-date information on the number and location of people in need of support.

Traditional demographic data sources, such as population and housing censuses, household surveys and administrative records, provide valuable demographic insights but are intrinsically static, offering only periodic snapshots of population distribution. These datasets fail to capture short-term fluctuations and spatial mobility, limiting their usefulness in crisis situations. In contrast, digital data sources such as social media and mobile phone data have the potential to provide dynamic pictures of the population, with temporal and spatial resolution limits often imposed only by privacy concerns\cite{Maas2019}. Several data sources have been investigated for this purpose in recent years, from mobile phone call detail records \cite{Deville2014, Lai2019}, web presence \cite{Zagheni2012, Alexander2022, Zagheni2017, Katz2024}, social media activity\cite{Sinnott2017, Yildiz2017}, and satellite nightlight imagery\cite{Rogers2023}.

Despite their potential, these data sources shed light only on the limited and skewed subset of populations who have access to and use digital technologies, introducing biases that must be carefully accounted for. For this reason, extensive efforts have been dedicated to analysing the representativeness of these datasets and developing methodologies to enable their use in estimating overall population stocks\cite{Ribeiro2020, Grow2022, Hsiao2024, Mu2024}. This study extends these efforts by investigating user location information, a novel social media-derived data source that enables the observation and tracking of population distributions and interactions. In particular, here we make use of spatially and temporally resolved Facebook user counts, that is, the number of active platform users within a given area and time period who agreed to share their locations\cite{Maas2019}. 

Previous research has demonstrated the utility of these data for monitoring mobility trends and population displacement during crisis events\cite{Shepherd2021, Gonzalez2022, Rowe2022}. However, to establish if and how they can be used to obtain reliable estimates of absolute population numbers, it is necessary to investigate the relationship between the measured user counts and known population stocks. By modelling how the connection between the observed social media numbers and the true underlying population varies across different socio-economic and geographical contexts, we can both better understand the drivers behind the differences in the technology uptake and predict how this might change when the characteristics of a given region do.

For our analysis, we focus on the case study of the Philippines for several practical and methodological reasons. First, the Philippines is one of the countries with the greatest exposure to natural disasters, including typhoons, earthquakes, volcanic eruptions and flooding, which require extensive humanitarian interventions\cite{Disasters2015}. However, the country currently lacks a fully structured disaster response infrastructure\cite{PhilippinesDisasterResponse2023}, making dynamic and near real-time population estimation particularly valuable. By focusing on the Philippines, therefore, we demonstrate how a country could take advantage of social media data and a dynamic approach to population stock estimation, with potential applications in other regions facing similar disruptions. Second, the Philippines provides an ideal testing ground due to its relatively high Facebook adoption level, estimated at 90.8 million users or $\sim$78\% of the population in 2025\cite{DataReportal2025}. This widespread platform use enables, for instance, robust analyses of population's vulnerability and exposure to climate shocks\cite{Dujardin2025}. Finally, we limit our scope to a single country because to develop an effective humanitarian response, it is important to have access to population estimation at the finest possible spatial resolution. Producing such estimates requires an analysis that considers hundreds or thousands of administrative regions within each country. Given the associated computational demands of such an approach, a geographically constrained case study is both practical and necessary. 

\begin{figure*}[!h]
    \centering
    \includegraphics[width=1\textwidth]
    {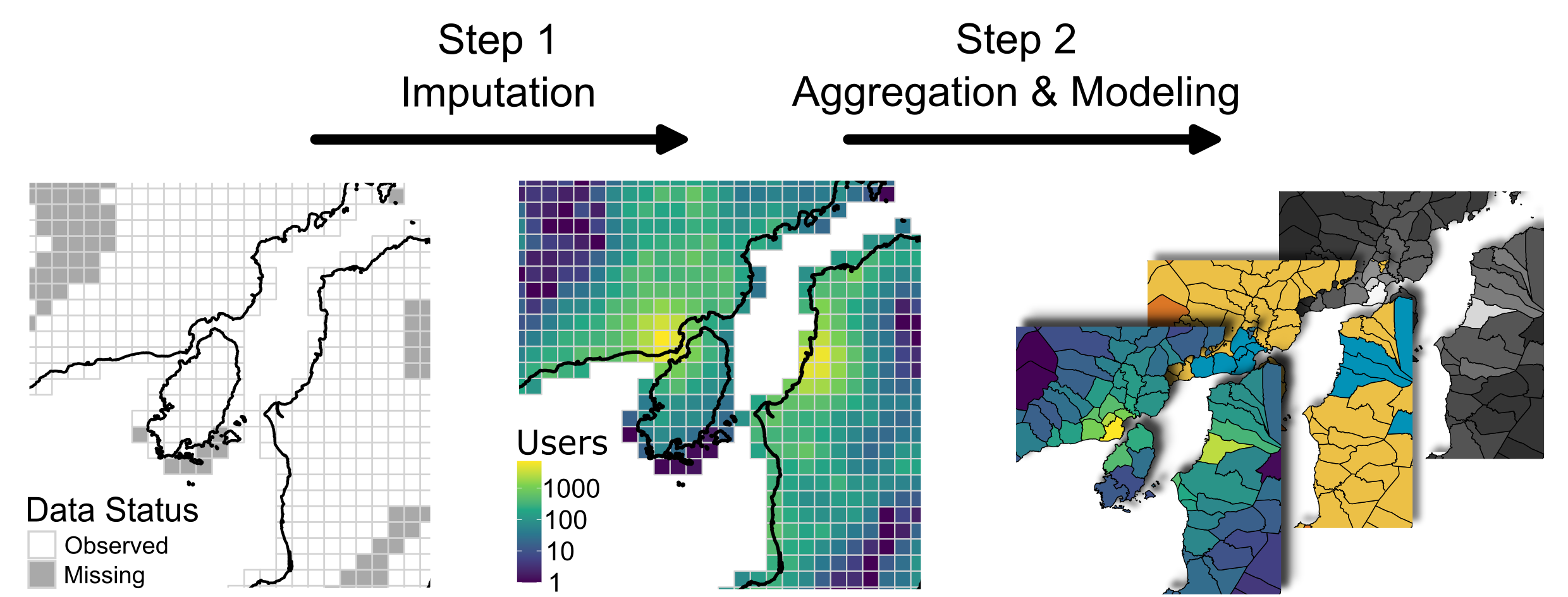}
    \caption{Illustration of the analysis workflow. The first step targets the missingness in the Facebook data at the tile level. A Bayesian statistical approach is used to impute the user count at the locations for which figures are not reported at the time point of interest (4 May 2020). Secondly, the imputed data is aggregated at the administrative area level and the proportion of facebook users is calculated using existing census data. This proportion is the target of a second model that includes a series of geolocated predictors (pictured on the right end side as stacked choropleth maps) that are used to explained the observed social media uptake.}
    \label{fig:workflow}
\end{figure*}

We build the analysis workflow following a Bayesian framework, starting from systematically addressing the issue of missingness in the high spatial resolution Facebook user count data (see Fig. \ref{fig:workflow} for a schematic representation of the workflow). These missing values arise due to the platform's differential privacy mechanisms, which introduce data censoring processes to preserve user privacy (see Section \ref{sec:imputation} for further details). Despite the widespread use of Facebook data in population and mobility research, few studies explicitly model this privacy-driven missingness; existing approaches typically rely on ad hoc smoothing or simple interpolation strategies that do not fully account for the uncertainty in the underlying censoring process\cite{Duan2024}. Secondly, we aggregate the enhanced counts at the municipality level and use census population figures, degree of urbanisation, and nighttime radiance data to develop a model of the Facebook uptake rate in each area. These rates are the scale factors used to convert observed social media user to population counts. Finally, we use the fitted model to predict the Facebook adoption rates of out-of-sample locations, demonstrating the possibility to forecast future population dynamics, which can be used to support more targeted and efficient responses to humanitarian crises.

In the following, we provide a detailed overview of the datasets used in this work, discussing their sources, limitations, and the preprocessing steps necessary to integrate information from multiple data streams (Section \ref{sec:data}). Following this, in Section \ref{sec:imputation}, we present the imputation model developed to account for the missingness in the Facebook count data. Section \ref{sec:data_exploration} provides a brief exploration of the variables considered in this work and their relationships, while Section \ref{sec:models} outlines the statistical methodologies employed to model the conversion rates between Facebook user counts and population stocks. In Section \ref{sec:discussion}, we interpret the model results and Section \ref{sec:limitations} outlines the potentials and limitations of our approach. Finally, Section \ref{sec:conclusions} summarises the main contributions of this work and suggests possible future research directions.

\section{
    \label{sec:data}
    Data Sources
}

A wide range of variables are expected to influence the rate of information and communication technologies usage within a population. Socioeconomic factors play an important role in determining who has access to these technologies and who has acquired the skills to operate them\cite{Gutierrez2010}. Demographic features such as age, gender, cultural and religious affiliation, and immigration status can also contribute, with their impact depending on the specific country or region\cite{Haight2014, Charness2020, Basson2010}. In addition, when analysing the Facebook user density spatial distribution, several other aspects might need to be considered; for example, the existence of inhabitable areas or the proximity to infrastructure, such as airports, can affect the observed user numbers.

To develop a model that is both practical to implement and can be extended to other locations and time periods, we incorporate several openly accessible data sources that provide contextual information for interpreting Facebook user counts alongside census-based population figures. These include the proportion of working-age inhabitants, the degree of urbanisation, nighttime satellite radiance measurements, and internet network performance. These factors can be interpreted as proxies of socio-economic indicators and, with the exception of the proportion of working-age inhabitants, can be frequently updated or calculated, potentially enabling near-real-time small area population estimations (SAPEs) as discussed below. Extensions to more complex models could be straightforwardly designed when additional predictors are available and depending on the required spatial and temporal resolution. For technical details concerning the data processing steps described below, please refer to the Supplementary Code Repository.

\subsection{
    \label{sec:data:boundaries}
    Philippines Administrative Boundaries
}

To integrate and harmonise all data sources geographically, we use the Philippines Subnational Administrative Boundaries, sourced by the Philippines Statistical Authority and the National Mapping and Resource Information Authority, and curated by the United Nations Office for the Coordination of Humanitarian Affairs (OCHA)\cite{PHLBound}. These boundaries provide geographical areas down to the barangay (administrative level 4) level, the smallest formal administrative division in the Philippines, and match those used for the census data collection. In this work, we aggregate the data at the administrative level 3 (Bayan or municipality) to balance the need for finely detailed estimates and a spatial resolution compatible with the other available data sources.

\subsection{
    \label{sec:data:census}
    Philippines Census Data and Working-Age Population
}

The most recent Philippines census data available at the beginning of our analysis was completed in 2020\cite{Mapa2021} and data at the barangay level are openly available on the OCHA data platform\cite{PHLCensus}. Due to the emergence of the COVID-19 pandemic and the stringent travel restrictions consequently implemented to contain the spread of the virus, the census data collection scheduled to occur in May was delayed to September\cite{Vera2020}. However, the surveys were still designed to reflect the population as of 1 May 2020. That the timestamp of the reference population falls within the COVID-19 lockdown period can have important consequences, as highlighted in Section \ref{sec:data:facebook}.

The census provides extensive demographic information, including population counts by gender and 1-year age brackets. This constitutes a rich picture of the context for each community and can be used to extract several summary statistics, such as age and gender distributions. Following other recent work in the field\cite{Mu2024}, we select to use the working-age proportion of the population (inhabitants between 15 and 65 years of age) as a demographic feature derived from the census data. We exclude other age-related variables to avoid collinearities across the predictors, which could complicate model fitting and interpretation, and because many demographic indicators are updated infrequently, limiting their usefulness in real-time applications. Rather, we treat the working-age proportion as a socio-economic measure that could be replaced by other, easier to collect, indicators of an area’s productivity (e.g. building height satellite data\cite{Pesaresi2023} or real-time transit data\cite{MTA}), allowing for the calculation of near-real time SAPEs.

\subsection{
    \label{sec:data:DUC}
    Degree of Urbanisation Classification
}

We use the degree of urbanisation classification (DUC) for 2020 as an additional proxy for the socio-economic status of a geographical area. This classification is based on the definitions of urban, peri-urban, and rural areas endorsed by the United Nations Statistical Commission as the methodology for international comparisons. Following the procedure developed by the Statistical Office of the European Union\cite{SMOD}, we start from the 1 km² data grids for settlement layers\cite{Schiavina2023} and population\cite{Carioli2023}, and map them onto the administrative units used in this work. These datasets are projected into the World Geodetic Coordinate Reference System (WGS 1984) to simplify the interoperability with other variables, resulting in a 30-arcsec-resolution grids that we use for subsequent analyses\cite{Maffeini2023}.

The settlement layers provide a DUC calculated by combining land/buildings area coverage with gridded population density values\cite{SMOD}. In the context of the model developed here, this should not be intended as the real-time population counts (the target of our modelling efforts), but rather as long-term estimated residency numbers. To project the DUC information from its 30-arcsec resolution to the municipality level, we first calculate the overlap between each grid cell and the administrative boundaries. The areas of the overlapping regions are then used to calculate the number of people that each degree of urbanisation class contributes to each Bayan, under the assumption that inhabitants are uniformly distributed over the cells. Finally, the class with the highest number of inhabitants is assigned to the whole administrative area. 

\subsection{
    \label{sec:data:nighttime}
    Nighttime Light Intensity
}

Nighttime satellite imagery provides additional information on the economic and technological activity of a region, traits that are commonly correlated with social media adoption, as discussed above. Here, we use monthly cloud-free Day/Night Band estimates generated by the Earth Observatory Group\cite{Elvidge2017}, accessed via the Payne Institute, Earth Observation Group portal\cite{VIIRS}. More specifically, we extract the average radiance values for May 2020, calculated after removing data impacted by stray light, over a 15-arcsec grid (approximately 500-by-500 meters at the equator). Each Bayan is then assigned a value equal to the mean radiance of the grid cells intersecting with its boundaries. While the information provided by this variable partly overlaps with the DUC, it still contains important insights towards understanding the distribution of Facebook users as our model will highlight. One clear appeal of using the nighttime light intensity is the possibility of frequent updates that could enable rapid refinements of the population estimates.

\subsection{
    \label{sec:data:devices}
    Network Usage
}
The last covariate we consider is the density of devices that initiated an internet network performance test in a given Bayan. Ookla's open data program\cite{OOKLA} provides this information quarterly updated for all 18 arcsec tiles (approximately 610-by-610 meter cells at the equator) worldwide. In particular, we use Ookla's API to extract the number of devices that used Speedtest\textsuperscript{\textregistered} by Ookla applications for Android and iOS in the second quarter of 2020 and map this data to the respective administrative areas. Each tile intersecting with a Bayan is weighed by the area of its overlap before calculating the total number of devices. Ookla provides separate numbers for devices using mobile and fixed networks but in the interest of simplifying our analysis, we combine these in a single network usage metric. 

\subsection{
    \label{sec:data:facebook}
    Facebook User Counts
}
The target variable in our study is the proportion of Facebook users in each Bayan. We obtain the user counts from a dataset sourced through Meta's AI for Good programme, which provides anonymised and aggregated statistics on Facebook app users who have opted to share the location history of their mobile devices\cite{Maas2019}. These figures are released by Meta for periods leading up to and including the response to humanitarian crises such as disease outbreaks and floodings. The dataset used in this work was collected by Meta as part of its efforts to support the response to the COVID-19 pandemic and Typhoon Vamco, which struck the Philippines on November 8th, 2020.

Meta assigns users to Bing tiles at level 13 (approximately 4.8x4.8 kilometres at the equator) during each of three daily 8-hour time windows, starting from midnight in Coordinated Universal Time (08:00 in Philippines local time). When users are recorded in multiple tiles during a time window, they are tagged in the location where they appear most frequently. The user counts are then aggregated for each tile and each time window. To integrate these figures with all other data sources, we assign Facebook user counts to the administrative units based on the geographical overlap between tiles and administrative boundaries, assuming that users are homogeneously distributed within each tile. For coastal regions, where tiles partially cover uninhabited areas, the population is first fully reapportioned to the land-based areas. 

\begin{figure*}[t!]
    \centering
    \subfloat[]
    {
        \label{fig:fb_over_time}
        \includegraphics[height=0.3\textheight]
        {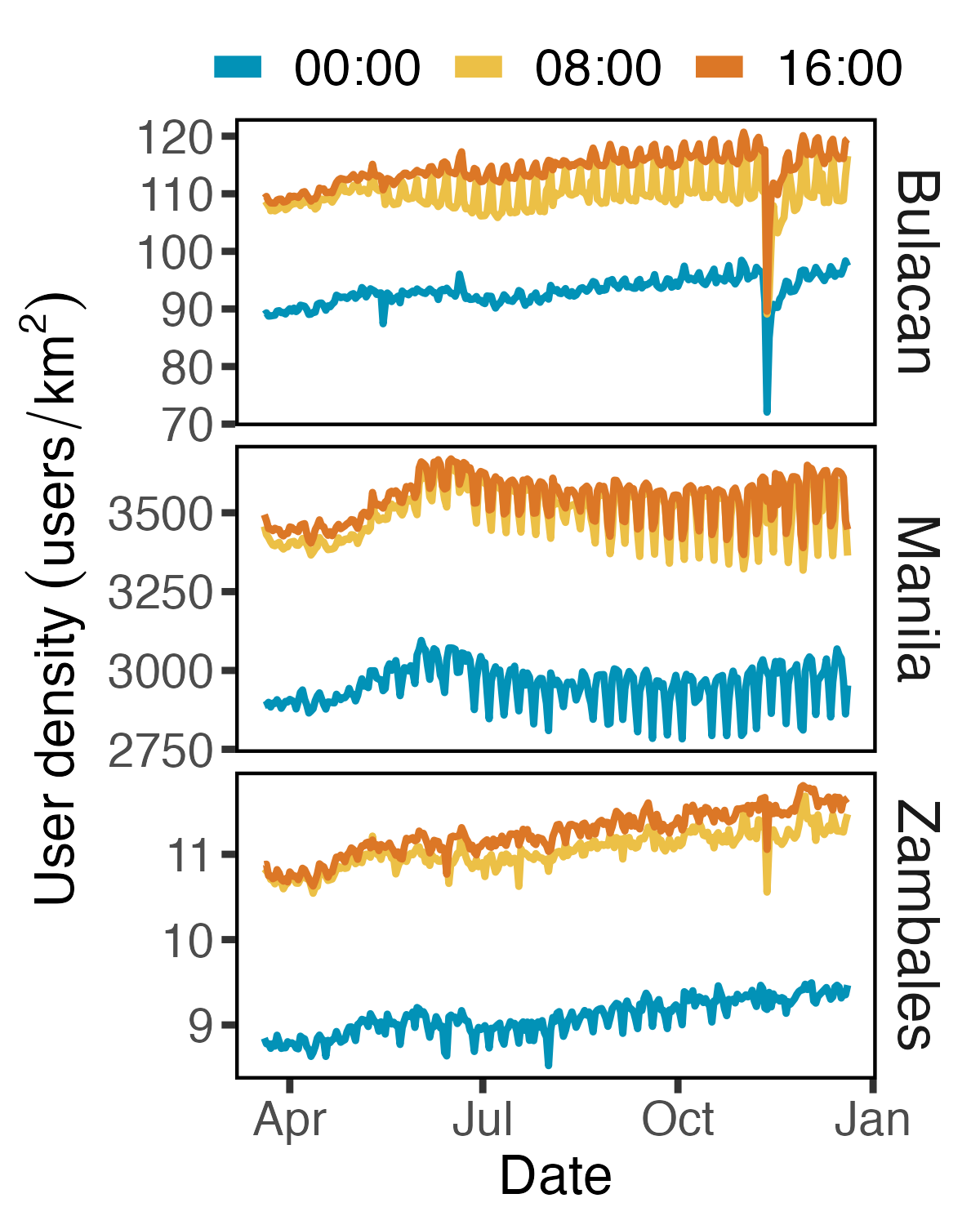}
    }
    \hspace{0.2cm}
    \subfloat[]{
        \label{fig:fb_weekdays}
        \includegraphics[height=0.3\textheight]
        {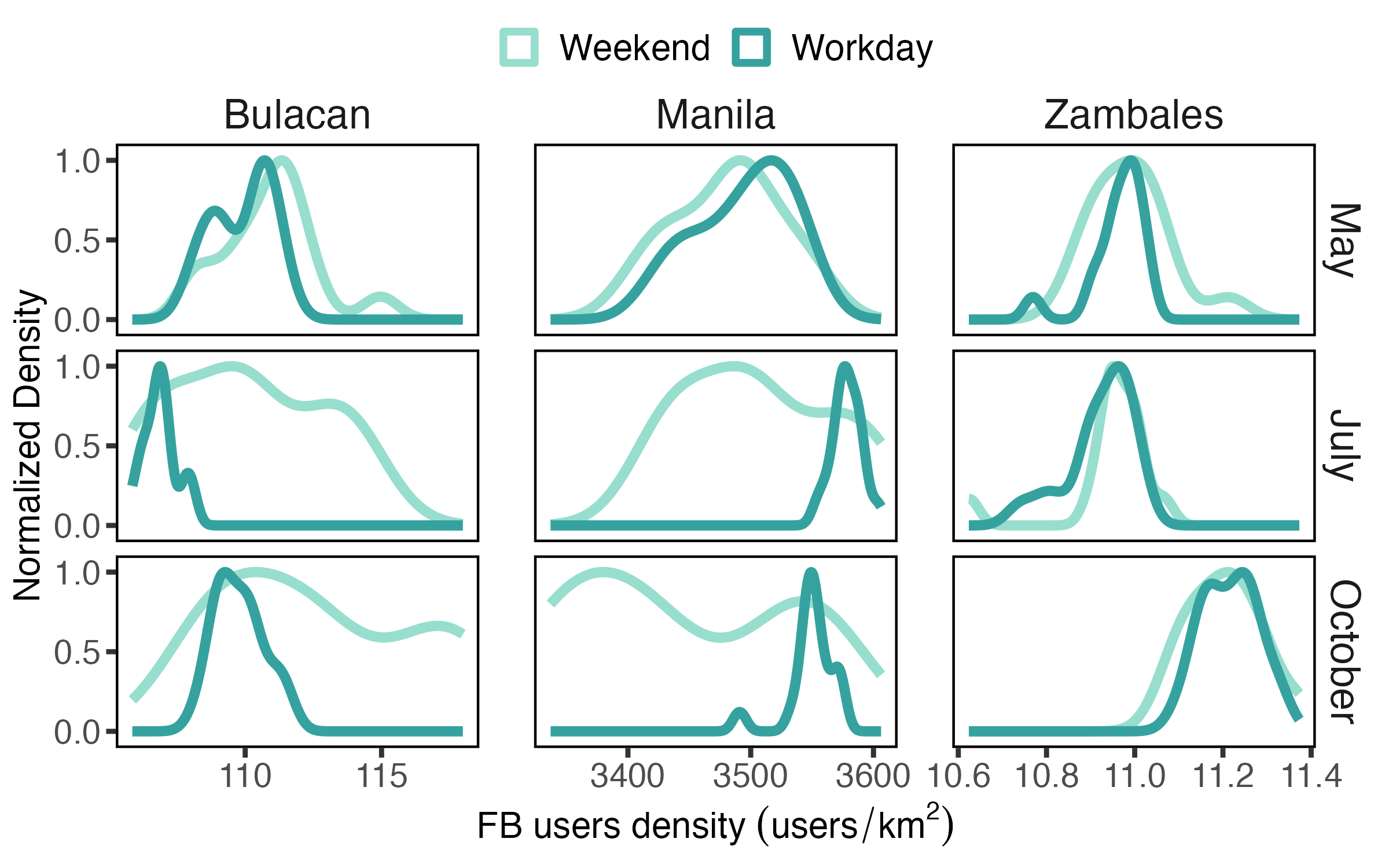}
    }
    \caption{(a) Facebook user density in three different municipalities showing different temporal trends in March-December, 2020. Data for the available timestamps (00:00, 08:00, 16:00) are plotted separately to highlight daily differences. The sudden drop in user density in November 2020 in Bulacan (and to a lesser extent in Zambales) reflects the effects of population evacuation in response to Typhoon Vamco. (b) Normalised region-level user density distribution for weekday and weekend time windows. Different user behaviour is clearly present in different regions, with Bulacan and the Manila areas witnessing respectively an increase and a decrease in user counts during the weekend (July and October data). On the other hand, user counts in Zambales are much less variable, reflecting its rural nature. Interestingly, in the month of May, the week-to-weekend shifts are strongly dampened. This reflects the impact of the restricted mobility imposed by the COVID-19 lockdowns, and suggests that the observed user counts can be closely matched against the census results (which reflect a person’s place of residence).}
    \label{fig:fb}
\end{figure*}

In Figure \ref{fig:fb_over_time}, we show the observed user density over a period in 2020 for three different Bayans and for the three time windows (all times are converted to Philippines local time). In Figure \ref{fig:fb_weekdays}, we instead plot the average weekday and weekend user density distribution for the three Bayans across three months. Both figures highlight a significantly different behaviour for the months of April and May (particularly in the larger Bayans of Bulacan and Manila), when users appeared to have drastically curtailed movements, as expected in the midst of one of the COVID-19 lockdowns. These observations informed the selection of a time window to use for the comparison between Facebook user counts and census data. In particular, selecting a weekday nighttime slot at the beginning of May 2020 ensures that people were most likely to be at their residence, the location where they are logged in the census data. We randomly select 4 May 2020 as the reference date for our analysis, ensuring an ideal alignment with census population figures.

Several methodological challenges need to be considered when using Facebook data to infer population stocks. First, the update frequency for the users' location history heavily depends on their active smartphone usage, resulting in the clear drop in user density observed for the overnight time window, when fewer devices are transmitting location data. This drop appears to be approximately homogeneous across Bayans, with the nighttime signal being approximately 80\% of the daytime one. This suggests that the signal drop is likely attributable to the technical details of how the signal is recorded rather than to real differences in populations, as evidenced by the substantially smaller size of the weekly oscillations in the user density for each time window, which describe actual changes in users' locations. For this reason, we focus on developing a conversion model for a single time window; night-day scaling factors can then be used to convert detected Facebook user counts into population estimates.

Secondly, Meta applies several data processing steps to ensure the protection of the users' privacy before making the counts available. Firstly, user figures are mixed with those of neighbouring tiles through a spatial smoothing process. This step can have a significant impact on analyses that attempt to model spatial correlation dynamics, as the true latent patterns are partially masked. We note that tiles with no overlap with land areas can acquire an artificial signal through this mixing mechanism, and we remove them from our analysis. Additionally, tiles with counts lower than 10 for a given time window have their signal censored. This can have a sizable impact, particularly for tiles located in large and sparsely populated administrative areas, whose user numbers could be severely and disproportionately underestimated. We will investigate this problem in detail and develop a method to address it in Section \ref{sec:imputation}.

Another challenge arises from the mismatch between Facebook data tiles and administrative boundaries. Many municipalities are considerably smaller than a single Facebook tile, particularly in densely populated urban areas. As discussed above, when redistributing Facebook user counts from tiles to Bayans, we assume that users are spatially homogeneously distributed, an approximation that is likely not satisfied by tiles that include areas with high spatial population variability. A potential improvement would be to use high spatial resolution population estimates to apportion Facebook users based on these pre-existing numbers. However, the user-to-population conversion factor is precisely the target of our model, and no hypotheses about this relationship can be included as part of the data preparation process.  

Finally, our approach relies on the important assumption that our results, obtained here at a given point in time, continue to accurately represent the underlying population dynamics in the future (or more accurately, until the data from the next population and housing census are available). While considering a point in time when movements were severely restricted has advantages when comparing the Facebook user data with the true population counts, it likely captures a demographic snapshot that cannot be exactly extrapolated to all times. This is true more broadly when daily dynamics that can affect the demographic composition of a neighbourhood (commuting, schooling, shopping, etc.) are present. In addition, the average rate of adoption of smart devices and Meta's applications could change through time due to increased digital access, churn in user location sharing, and cohort replacement mechanisms. This would result in shifts in the user-to-population conversion factors. A hint that this change might be occurring is in the upward drift of the average user densities observed throughout 2020 (see Fig. \ref{fig:fb_over_time}), but this should be compared against the expected increase in population. Interestingly, data for the year 2021 show instead a widespread decreasing drift in user counts (Fig. \ref{fig:suppFB_2021}). The addition of a forecasting model for population growth and for technology adoption would provide a path to improving the accuracy of future population stock estimates; however, this is beyond the scope of our current work. 

Most of the challenges highlighted above originate from the intrinsic properties of the available data and need to be embraced in the model in the form of the assumptions that we introduced. Nonetheless, in the next section we introduce a statistical approach to address the effect of user number censoring. 

\section{
    \label{sec:imputation}
    Missing Facebook Data Imputation
}

\subsection{
    \label{sec:imputation:data}
    Approach
}

As mentioned above, the censoring process used by Meta to guarantee the users' privacy can potentially have an important impact on the measured user counts and, therefore, on the users-to-population conversion factors. 

\begin{figure*}[t!]
    \centering
    \subfloat[]
    {
        \label{fig:imputation_hist}
        \includegraphics[width=0.45\textwidth]
        {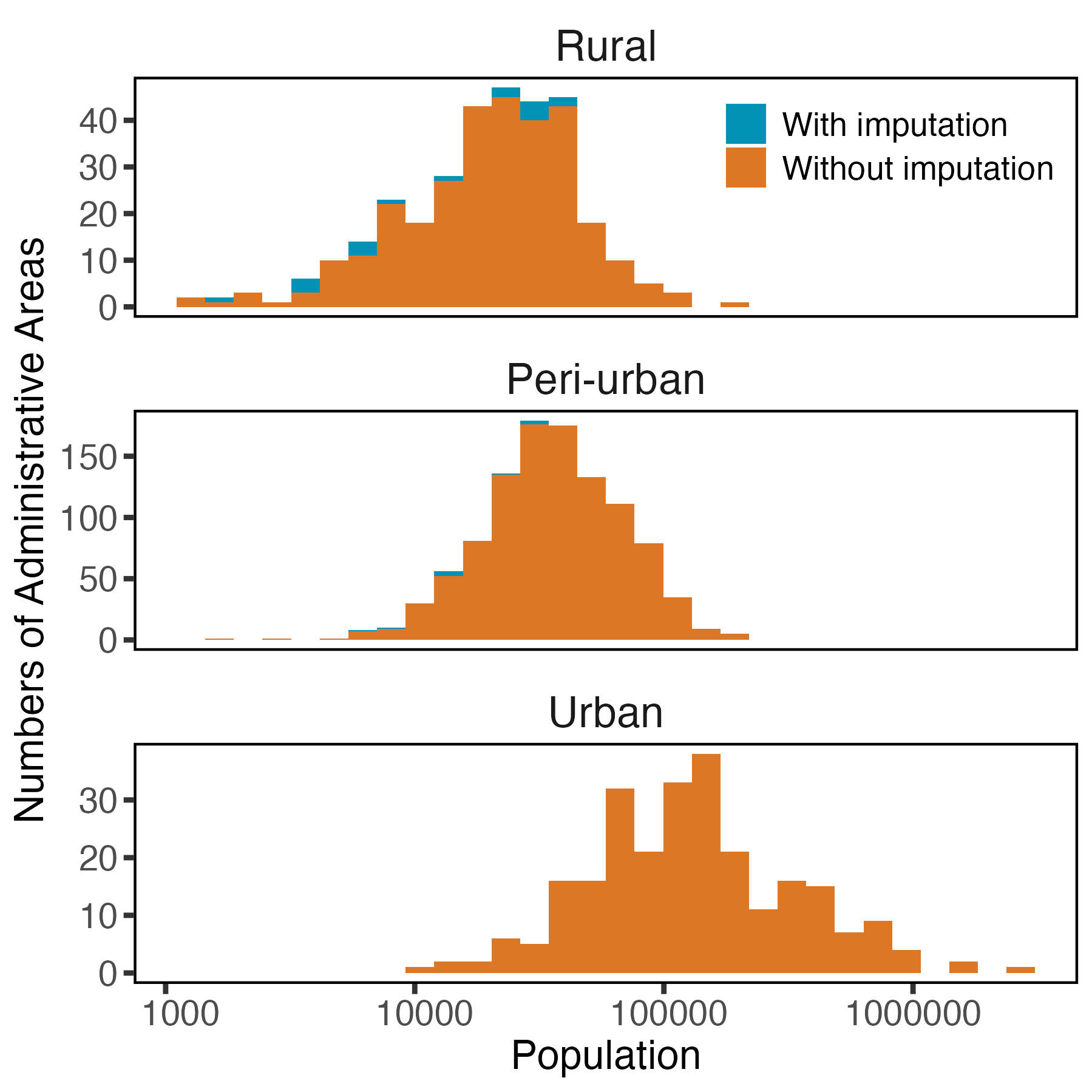}
    }
    \hspace{0.2cm}
    \subfloat[]{
        \label{fig:imputation_scatter}
        \includegraphics[width=0.45\textwidth]
        {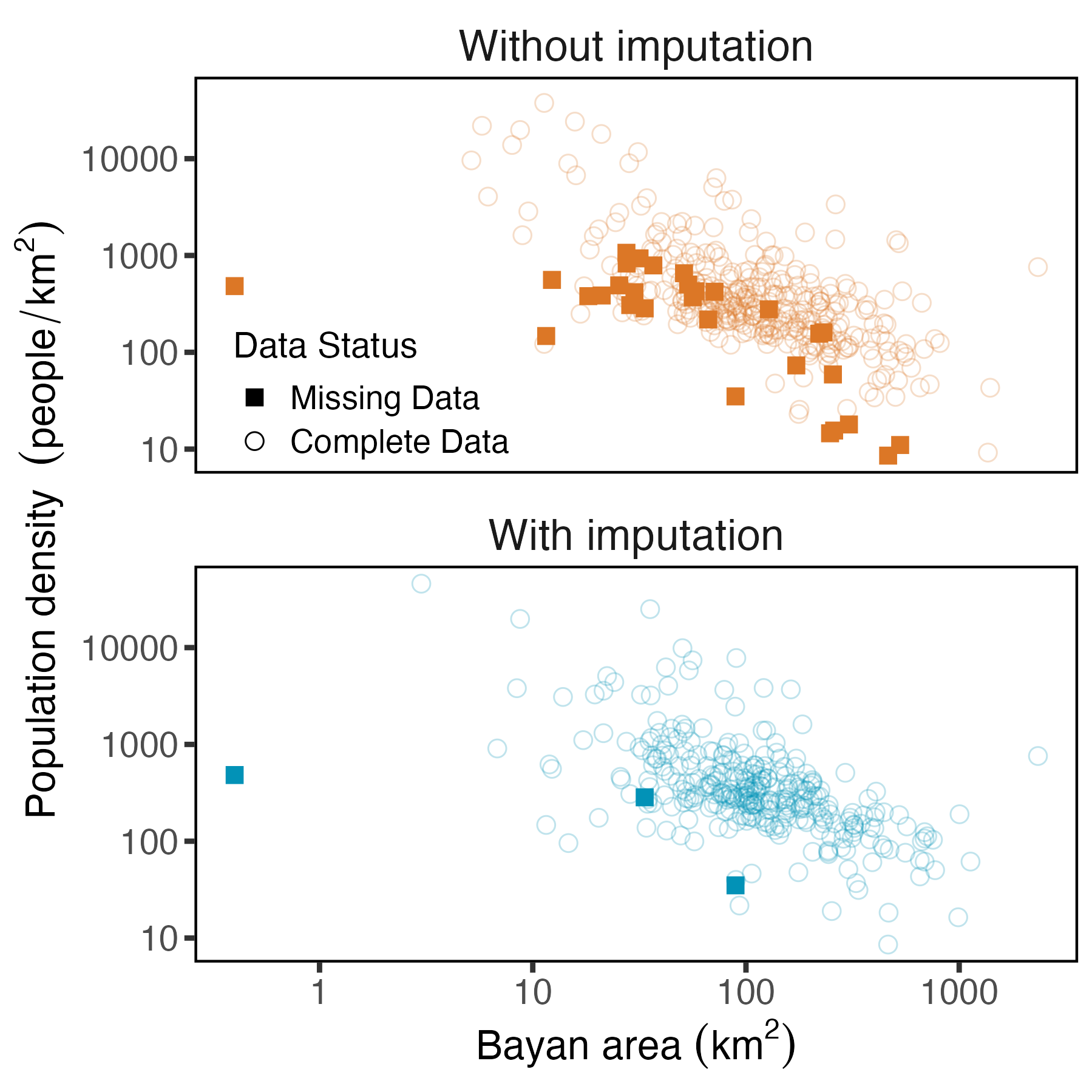}
    }
    \caption{(a) Histograms of all administrative areas by population size calculated before (orange) and after (blue) the imputation process is implemented. The three panels refer to the different degrees of urbanisation. Most of the areas recovered through the imputation process are Rural. (b) Scatter plots of the population density against the area of each administrative unit, with the markers' shape distinguishing between units with (circles) and without (squares) signal. Only a sample of the areas with signal is mapped for ease of visualisation. The two panels show the situation before (top) and after (bottom) the imputation process is implemented. Data missingness is biased towards small or sparsely populated areas.}
    \label{fig:imputation}
\end{figure*}

To estimate this impact, we select the signal for our timeframe of interest (nighttime window for 4 May 2020) and first remove the tiles with missing values that are fully associated with the DUC class 10 (a subset of the rural tiles identifying uninhabited areas and water bodies). Of the remaining tiles, approximately 37\% have a censored signal. When aggregated by administrative area, this translates to 1.7\% of Bayans with no counts and $>6.6\%$ of Bayans with a percentage of tiles with missing values higher than 80\%. More importantly, the censoring overwhelmingly affects regions that are rural and have low population density, as expected and as shown in Figure \ref{fig:imputation}.

In Figure \ref{fig:imputation_hist}, we compare the population size distributions of the administrative areas with an existing user count before and after the imputation process is applied. It is evident that rural areas are the most affected, particularly those with a smaller number of inhabitants.
Figure \ref{fig:imputation_scatter} sheds further light on this by highlighting the relationship between the characteristics of a given area and the presence or absence of user counts. We see that for a given Bayan's size, the regions lacking data coverage tend to have smaller population density. Both these insights point to a bias in the data missingness that we address in the attempt to reduce its effect on the model representativeness. In general, this type of bias should be considered when using social media censored data, given that several geographical and demographic characteristics can affect the censoring process. We note that, even after implementing the imputation step, three areas still have no user counts. These are small islands that fall outside of the Facebook grid coverage, as shown in Figure \ref{fig:supp_FB_miss_after_imputation}. 

\begin{figure*}[!h]
    \centering
    \subfloat[]
    {
        \label{fig:sample_dist}
        \includegraphics[height=0.27\textheight]
        {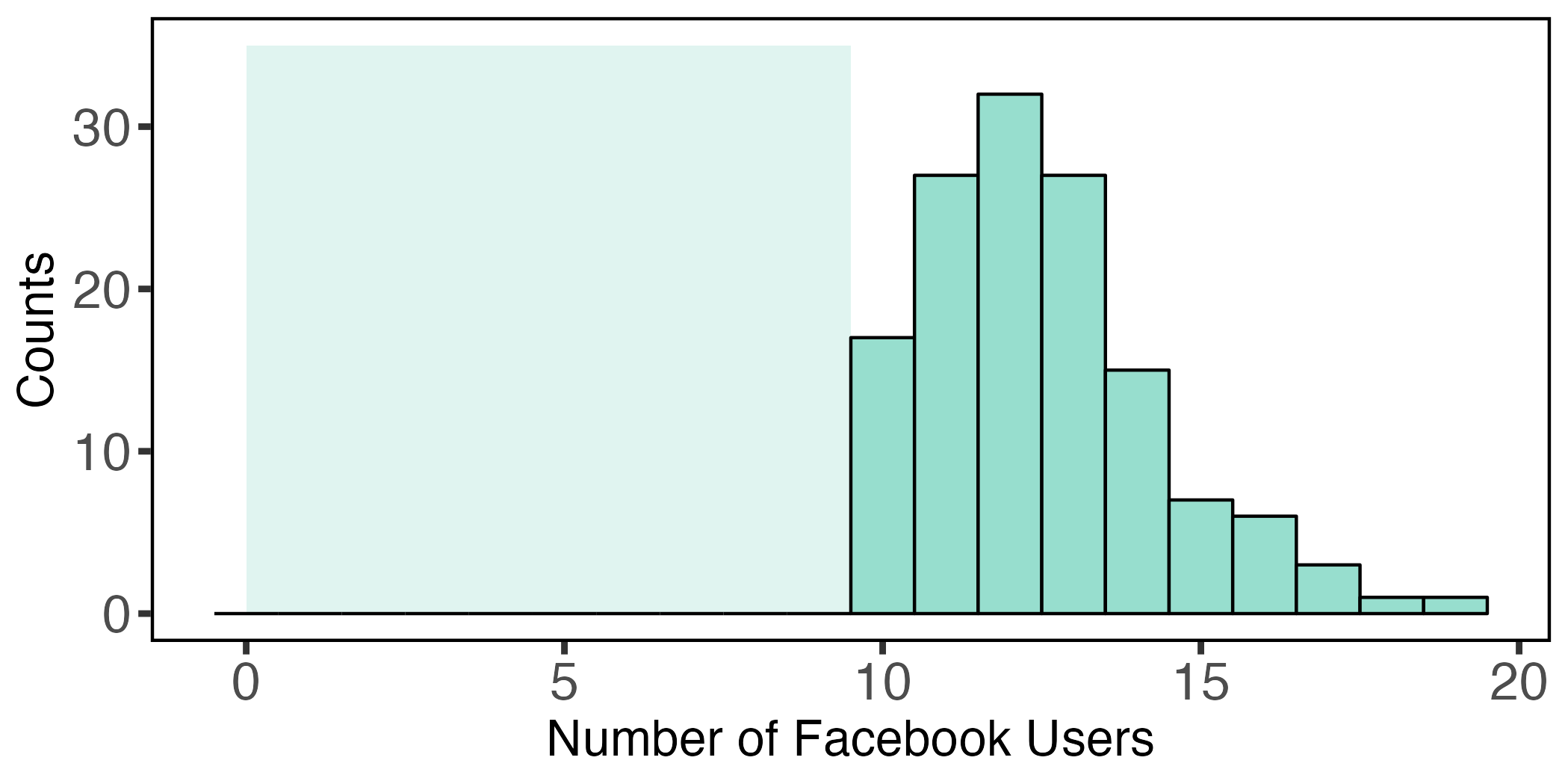}
    }
    \hspace{0.2cm}
    \subfloat[]
    {
        \label{fig:missing_dist}
        \includegraphics[height=0.27\textheight]
        {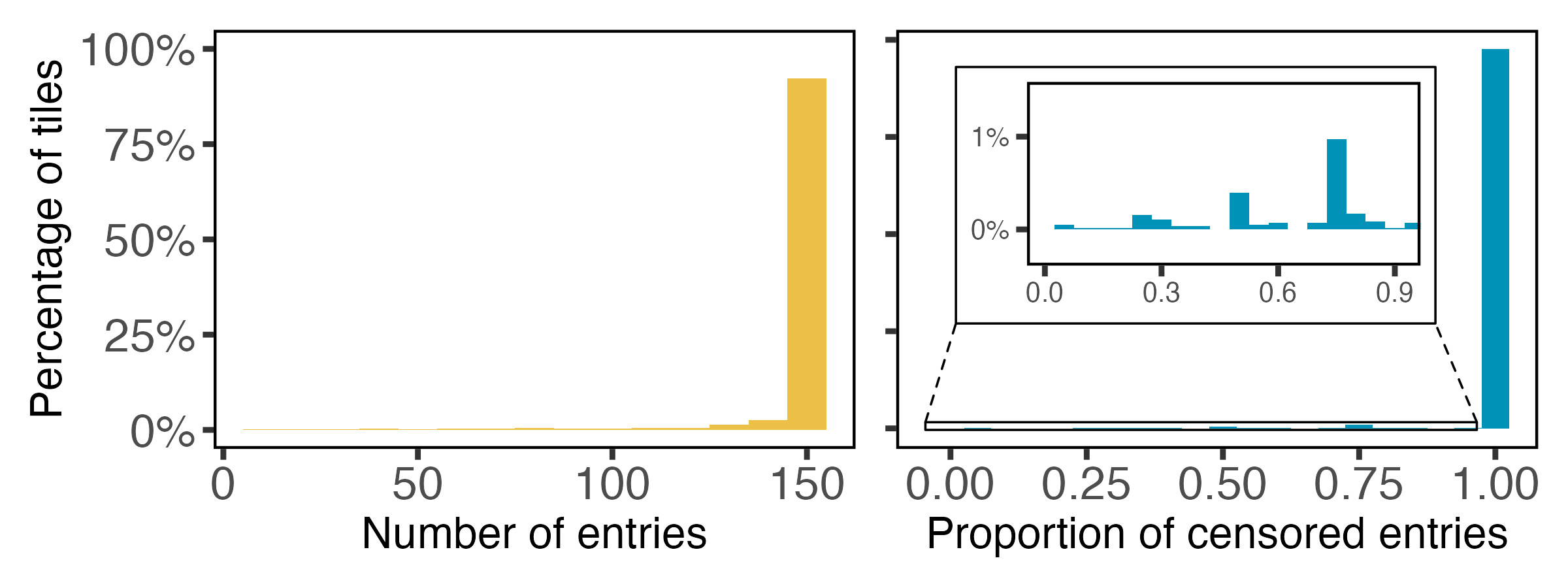}
    }
    \caption{Some statistical properties of the user counts for the tiles included in the imputation process (those with missing data on 4 May 2020). (a) Histogram of the number of Facebook users recorded throughout all the available 2020 weekly and nighttime data for a sample tile. The clearly truncated distribution highlights the effect of the censoring process. (b) Histogram (in percentage) of the number of times each tile occurs in the data. Most tiles have 151 entries. (c) Histogram of the proportion of censored entries per tile. For the large majority of the tiles used in the imputation process the signal is missing every day. Spikes in data availability are visible around 25\%, 50\%, and 75\% of censored entries.}
    \label{fig:missing}
\end{figure*}

The central idea behind the imputation approach is that missing user counts are informative rather than arbitrary. Counts are censored because they fall below a threshold, and the frequency with which this occurs for a given location conveys partial information about the missing values. To exploit this, it is crucial to consider the full distribution of user counts, including both observed and censored values, for each tile over time. Figure \ref{fig:sample_dist} illustrates this point by showing the observed count distribution for a single tile across several months, which is clearly truncated at the threshold of 10. Here we see that if one were to impute missing values using only the observed counts, for example by averaging them, the resulting estimates would be skewed upward because low-count observations are systematically excluded. Conversely, ignoring the full (observed and unobserved) count distribution and imputing missing values with a fixed heuristic such as the midpoint of the censored interval (5 in our case) would produce consistently downward-biased estimates. 

In line with these observations, data covering all the available time frame for the year 2020 (March to December) for all the locations with missing data on 4 May 2020 is included in our imputation approach. To ensure that the data included in our analysis are representative of the true underlying distribution for the reference date and time, we exclude signal from the daytime windows and from weekend days. We assume that all the remaining counts for a given tile are samples from the same distribution, implying that the data generating process does not change throughout the year. We note that we cannot take the other variables discussed above (including the geographic location of tiles) into consideration as part of the imputation process because their association with the user counts will be the target of the user-to-population conversion model. 

Figure \ref{fig:missing_dist} outlines some distributional characteristics for the tiles included in the imputation process. The majority of these tiles appear in the data for every day in the record ($N_{days} = 151$) (left panel); however, $\sim 13\%$ of them have fewer entries. In addition (right panel), a large proportion ($\sim 97\%$) of tiles have only censored values recorded in the dataset. For those with uncensored values (136), the proportions of available data seemingly peak at 25, 50 and 75\%. This, together with the incomplete coverage for some tiles, might point to data collection errors, which we expect can be addressed in the future as user tracking procedures improve. Figure \ref{fig:map_missing} shows the geographical location of all of the tiles with missing values on 4 May 2020, distinguishing those that only have censored values in the data from the others. 

We heavily relied on the observed properties of the missing data to design the imputation approach. First, we consider all the available entries for the tiles with at least one observed user count over the time period of interest. We group the remaining tiles (with all censored entries) according to the number of times they appear in the dataset. Because data frequency is the only information used by the imputation model, we consider all tiles in each of these groups to be statistically equivalent and a single reference location is used to infer the data distribution for all matching tiles. The underlying assumption of complete equivalence for all tiles in a group is not completely realistic, but it is justified in the context where no other distinguishing information is used. We note that the number of tiles with only censored values itself contains information on the expected user counts. For instance, if we assume that all tiles considered in the imputation process are part of a single "population" of low user density locations, the presence of more samples with no recorded values would imply a lower expected count for all locations in the context of a hierarchical model (where information is shared between tiles). However, a comparison between the results of hierarchical and non-hierarchical approaches (see Fig. \ref{fig:imp_ppc_non_hier}) reveals negligible differences, suggesting that the large number of data points for each tile might largely overwhelm the cross-tile information sharing effect even if all tiles were singularly considered.

This tile-grouping simplification is not strictly necessary, but it greatly reduces the computational complexity of the model. We note that, if resources are not a limitation, the imputation process described below can be combined directly with the users-to-population model described in Section \ref{sec:models} to obtain conversion factors for each of the almost 16,000 tiles (after appropriate disaggregation of the census data).

\subsection{
    \label{sec:imputation:model}
    Model Definition
}

To recover the parameters of the user distribution for each tile and infer the missing count numbers, we develop a Bayesian Poisson count hierarchical model using RStan, the \textbf{R} interface to the probabilistic programming language Stan (details for all packages used in our analysis are available through the code repository). 

The model structure is typical of censored data models \cite{StanCensored}, which include two different contributions to the likelihood of the reported counts $\mathcal{L}(u|\lambda_{t})$ depending on the observation status of a data point:
\begin{equation*}
    \mathcal{L}(u|\lambda_{t}) =
        \begin{cases}
            \text{Pois}(u_{obs}|\lambda_{t}) & \text{if data is observed,} \\
            F_{\text{Pois}}(u_{cens}<10|\lambda_{t}) & \text{otherwise.}
        \end{cases}
\end{equation*}
Here, $u_{obs}$ and $u_{cens}$ represent the observed and censored Facebook user counts, respectively, $\lambda_{t}$ is the expected population of tile $t$ and $F_{\text{Pois}}(y < k|\lambda_{t})$ is the Poisson cumulative distribution function, representing the likelihood of observing a number of counts lower than 10 given the rate parameter $\lambda_{t}$.

To complete the model, we add an exponential prior for each of the tile-dependent rate parameters and hyperpriors that controls the distribution of the $\lambda_{t}$ scale parameters:
\begin{align*}
    \lambda_{t}|s \sim \text{Exp}(s) \\
    s \sim \text{Exp}(5)
\end{align*}
where the exponential distribution is parametrised with its scale (inverse of the rate) to simplify the parameter interpretation.

Finally, the parameter space is explored using the No-U-Turn Sampler algorithm implemented in RStan to reconstruct the parameters' posterior distributions.
Inference is first tested on synthetic data, demonstrating that we can recover the true parameter values and confirming that the model is correctly specified (see Supplementary Information \ref{fig:imp_trace_plots_sim}, \ref{fig:imp_ppc_sim}). 
In Figure \ref{fig:imp_trace_plots}, we demonstrate that, when the model is applied to real data, the Markov chain Monte Carlo (MCMC) chains show a healthy sampling behaviour after warm-up, indicating convergence and appropriate mixing\cite{Roy2020}. 

\subsection{
    \label{sec:imputation:results}
    Model Results and Data Imputation
}

\begin{figure*}[!h]
    \centering
    \includegraphics[width=1\textwidth]
    {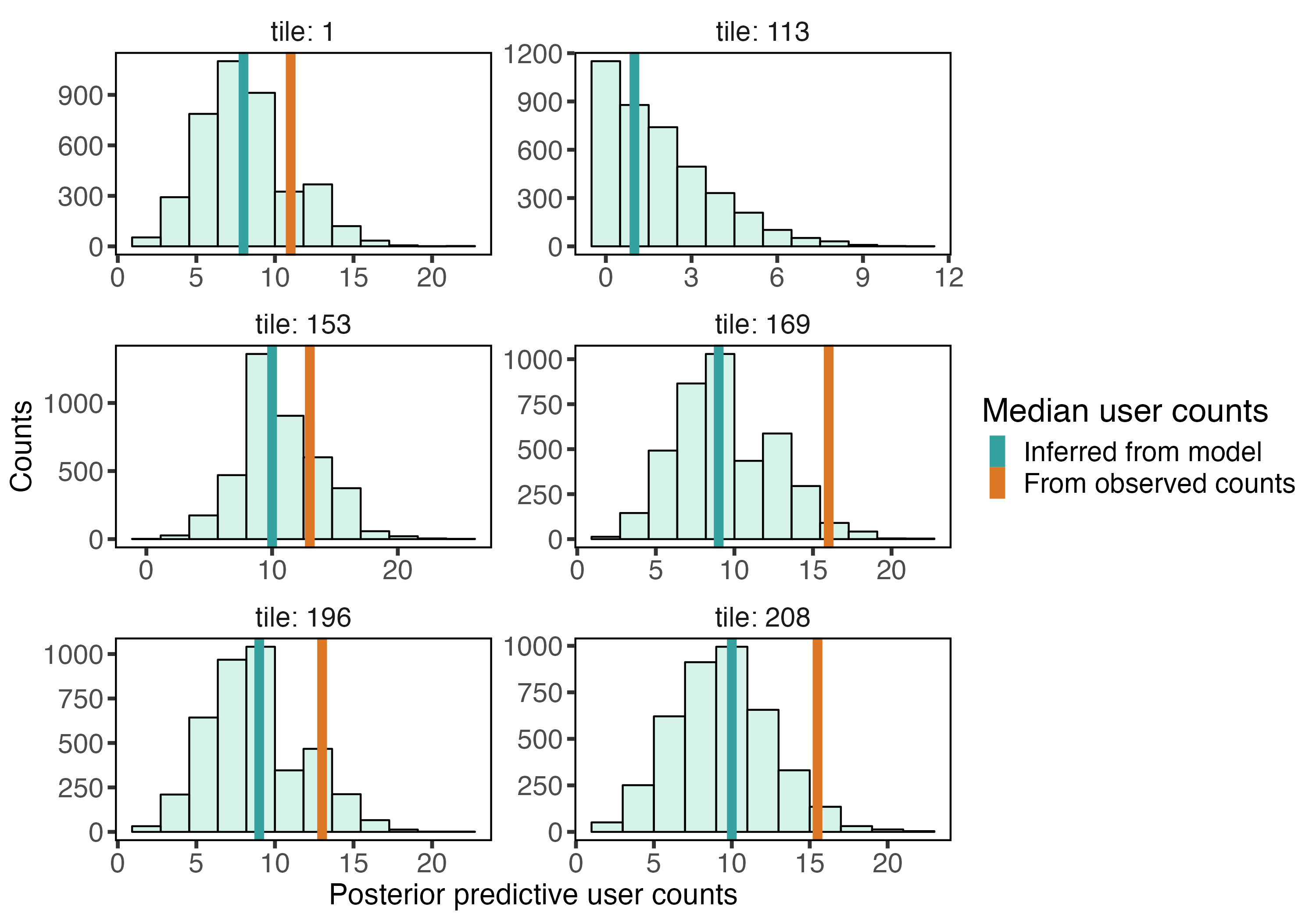}
    \caption{Histograms of user count predictions generated by the imputation model for six sample tiles. For all tiles with partially observed data, the model-inferred user counts median is lower than that of the observed values.}
    \label{fig:imp_ppc}
\end{figure*}

To assess the model performance, we first use draws from the posterior predictive distribution to compare against the observed data. The results for six tiles are collected in Figure \ref{fig:imp_ppc}, where we present the distribution and median of the model-simulated user counts and the median calculated using only the observed data.  Excluding tile 113, which records exclusively censored counts, we see that the median inferred from the imputation model is always lower than the empirical median, an expected consequence of the data censoring process, as explained above.  

Finally, we impute all the censored values for 4 May 2020 using the following procedure. For the tiles that were individually modelled, a sample from the respective posterior distribution is selected. To the remaining tiles, which only have censored values in the dataset, we assign a sample from the posterior distribution associated with their tile group, comprising tiles with the same number of entries in the data. In all cases, the samples are weighted by the percentage of inhabited land in each tile. This last step is based on the assumption that the population is uniformly distributed across the settled areas within each tile. This assumption, while not generally satisfied, is realistic for low density rural regions. 

Once each tile has been assigned a (potentially zero) user count, we can aggregate the number of users per Bayan as described above. Table \ref{table:imputation_results} presents descriptive statistics on the impact of the imputation model. Imputing missing values does not add a large absolute number of users, but it has a sizable effect on subsets of the data, particularly in low-density and rural areas. A robust imputation process, like the one we developed, is therefore important to obtain figures that are representative across all geographical locations.

\begin{table}[t!]
\centering
\begin{tabular}{|p{9cm}|C{3cm}|}
 \hline
 \multicolumn{2}{|c|}{\textbf{Increase in values post-imputation}} \\
 \hline\hline
 User counts  & 0.4\% (4675953)\\
 Bayans with signal  & 1.7\% (1636) \\
 Low population ($<10000$ people) Bayans with signal & 11.8\% (95) \\
 Rural Bayans with signal  &  5.5\% (323) \\
 \hline\hline
\end{tabular}
\caption{Percentage increase in available data after the implementation of the imputation process. The figures between parentheses are the final numbers of entries for each category after imputation.}
\label{table:imputation_results}
\end{table}

\section{
    \label{sec:data_exploration}
    Data Exploration
}

\begin{figure*}[!h]
    \centering
    \includegraphics[width=1\textwidth]
    {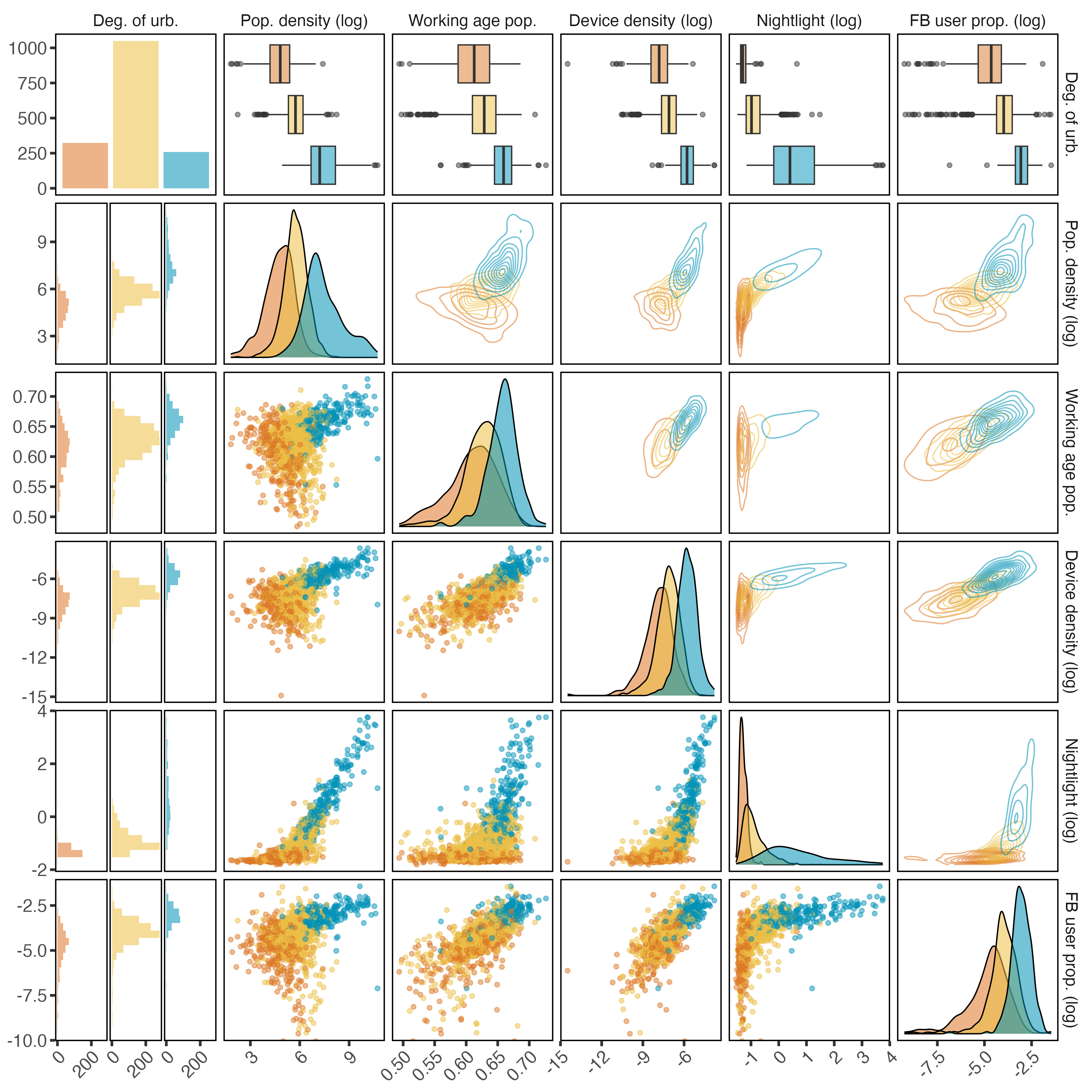}
    \caption{Summary visualisation for all the investigated variables. The colours reflect the different degrees of urbanisation (orange = rural, yellow = peri-urban, and blue = urban). Data with missing information is dropped in the device density plots. Population density, device density, nightlight intensity, and the proportion of Facebook users are mapped to a logarithmic scale.}
    \label{fig:pair_plot}
\end{figure*}

Figure \ref{fig:pair_plot} provides an overview of the pairwise relationship between the variables. 
The graphs in the last row demonstrate that all predictors are strongly associated with the proportion of Facebook users in a given area. More specifically, higher DUC, population density, proportion of people of working age, number of recorded devices, and nighttime intensity are related to higher proportions of Facebook users on aggregate. However, it is interesting to observe that, when disaggregated by DUC, higher population densities are associated to lower social media presence in rural areas (see second plot from the left in the last row). This might be the result of higher poverty levels in high density rural areas, which is also reflected by a decrease in the proportion of working age people (see third row). 

The correlations between the predictors are particularly obvious when inspecting the fourth row of graphs, which highlights a clear (and expected) association between the number of devices that initiated a network perfomance test and all other variables. Because of this strong correlation and the sizable number of Bayans lacking information on the number of devices, we do not include this variable in the models introduced below. However, the close relationship between the Ookla's data and the other predictors and its rapid accessibility suggest that, when the data completeness improves, it could act as an ideal proxy for population figures, potentially even as an alternative to the Facebook data.

We also exclude information on the population density calculated from the census data from the remainder of the analysis. While this is a strong predictor of the proportion of Facebook users, our goal is to develop models that can infer the true population figures from predictors readily available with high temporal resolution. 

The directed acyclic graph shown in Figure \ref{fig:dag} shows the possible causal linkages between all the variables we considered. The proposed structure has important consequences for the interpretation assigned to the model parameters, but this causal interpretation is beyond the scope of this current work. Here, we only use the graph to generate simulated data used to test the models. 

\section{
    \label{sec:models}
    Models to estimate population stocks
}

To ensure the robustness of our results and to shed light on the impact of different contributions, we progressively increase the model complexity and test it on the simulated data. All models are developed within a Bayesian framework and coded using RStan; for each model, we run the MCMC sampling algorithm on 4 parallel chains and 1000 iterations post warm-up. The models' target is the proportion of Facebook users in each administrative areas, from which the real-time population can be calculated using the measured user counts. We train the models on 80\% of the administrative areas for each DUC, and reserve the remaining 20\% to test their predictions (Fig. \ref{fig:train_test_split} shows the locations included in the train and test sets). We use the models' predictions for the test locations to measure how well they can extrapolate to new contexts. 

\subsection{
    \label{sec:model_bin}
    Binomial Model
}

As a starting point, we consider a hierarchical binomial count process model ($M_{bin}$). The likelihood of the data for each municipality is a Binomial distribution with probability of success (i.e. the probability of being a recorded Facebook user) dependent on the model predictors through the inverse logit link transformation.

\begin{align*}
    FB_{i} &\sim \text{Bin}(n = N_{i},\,p = p_{FB,i})  \\
    p_{FB, i} &= \text{logit}^{-1}(a[U_{i}] + b_{w}[U_{i}] \cdot W_{i} + b_{l}[U_{i}] \cdot \text{log}L_{i}) \\[8pt]
    \begin{bmatrix}
        a[U_{i}] \\
        b_{w}[U_{i}] \\
        b_{l}[U_{i}]
    \end{bmatrix}
    &\sim \text{Normal} \left\{ 
    \begin{bmatrix}
        a_{mu} \\
        0 \\
        0 
    \end{bmatrix} ,
    \begin{bmatrix}
        a_{sigma} & 0 & 0 \\
        0 & b_{w, sigma} & 0 \\
        0 & 0 & b_{l, sigma}
    \end{bmatrix}
    \right\} \\[8pt]
    a_{mu} &\sim \text{Normal}(-4,\,1)  \\
    a_{sigma} &\sim \text{Exp}(1)  \\
    b_{w, sigma} &\sim \text{Exp}(1)  \\
    b_{l, sigma} &\sim \text{Exp}(1)   
\end{align*}

Where the index $i$ identifies each Bayan and $FB_{i}$,  $N_{i}$ and $p_{FB, i}$ are, respectively, the number of recorded Facebook users, the number of inhabitants, and the probability that a person in the Bayan is a recorded Facebook user. $a[U_{i}]$, $b_{w}[U_{i}]$, $b_{l}[U_{i}]$ are the varying intercept and slopes, all modelled as effects dependent on the local DUC $U_{i}$ (Rural = 1, Peri-urban = 2, Urban = 3). $W_{i}$ and log$L_{i}$ represent the proportion of people of working age and the logarithm of the nightlight intensity, respectively. All model parameters are assigned normal prior distributions with parameters for each DUC in turn sampled from exponential hyperprior distributions.

The Markov Chain Monte Carlo (MCMC) traces of a subset of the model parameters and pairwise plots of the posterior samples are shown in Figures \ref{fig:traces_no_sp} and \ref{fig:pair_plot_no_sp}. Despite the presence of significant correlation between two pairs of parameters, the plots demonstrate a healthy sampling behaviour, confirmed by the summary inference statistics (Supplementary Information Table \ref{table:stats_model_bin}). However, the data generated from the posterior distribution (Fig. \ref{fig:dens_ovly_no_sp}) suggests the existence of misspecifications in the models. This is further corroborated through the investigation of the Pareto $k$ diagnostics obtained using Pareto smoothed importance-sampling (PSIS) leave-one-out cross-validation (LOO-CV) \cite{Vehtari2017, Vehtari2024}, which, for several data points, significantly differs from the behaviour expected for a well-specified model (Fig. \ref{fig:psis_diagnostics}). In particular, data from urban areas appears to be poorly replicated and suggests that a model allowing for greater dispersion is required to fully capture the underlying data generating process. 

\subsection{
    \label{sec:model_beta_bin}
    Model with Overdispersion
}

To obtain a more faithful representation of the data generating process, we implement a beta-binomial regression model ($M_{betabin}$), which relaxes the dispersion limits of the binomial distribution by introducing a random beta distributed binomial probability of success. 

\begin{align*}
    FB_{i} &\sim BetaBin(n = N_{i},\,\alpha = \alpha_{i},\,\beta = \beta_{i}) \\
    \alpha_{i} &= \overline{p_{FB,i}}(1-\rho[U_{i}])/\rho[U_{i}] \\
    \beta_{i} &= (1 - \overline{p_{FB,i}})(1 - \rho[U_{i}])/\rho[U_{i}]) \\
    \overline{p_{FB, i}} &= logit^{-1}(a[U_{i}] + b_{w}[U_{i}] \cdot W_{i} + b_{l}[U_{i}] \cdot logL_{i}) \\[8pt]
    \rho[U_{i}] &\sim Beta(1,\,3)\\  
\end{align*}

Here, $\overline{p_{FB, i}}$ can be interpreted as the average probability that an inhabitant or region $i$ is a recorded Facebook user and $\rho$ is the degree of urbanisation-dependent overdispersion parameter of the Beta-binomial distribution. For simplicity, we do not model $\rho$ hierarchically. Rather, we assign a common set of parameters identified from the direct inspection of prior predictive samples generated by the model and chosen to ensure that the prior predictive data covers all reasonable scenarios. We note that, the large number of data points available for all DUC levels significantly reduces the impact of choices in the definition of the prior distributions. All the other parameters are defined as for $M_{bin}$.

It is important to note that overdispersion likely hides the presence of additional complexities in the data generating process that our model cannot capture (i.e. factors encoding important distinctions between the urban Bayans). Nonetheless, an improved description of the real data dispersion provides more accurate uncertainty estimations.

\begin{figure*}[!h]
    \centering
    \subfloat[]
    {
        \label{fig:dens_ovly_no_sp}
        \includegraphics[height=0.34\textheight]
        {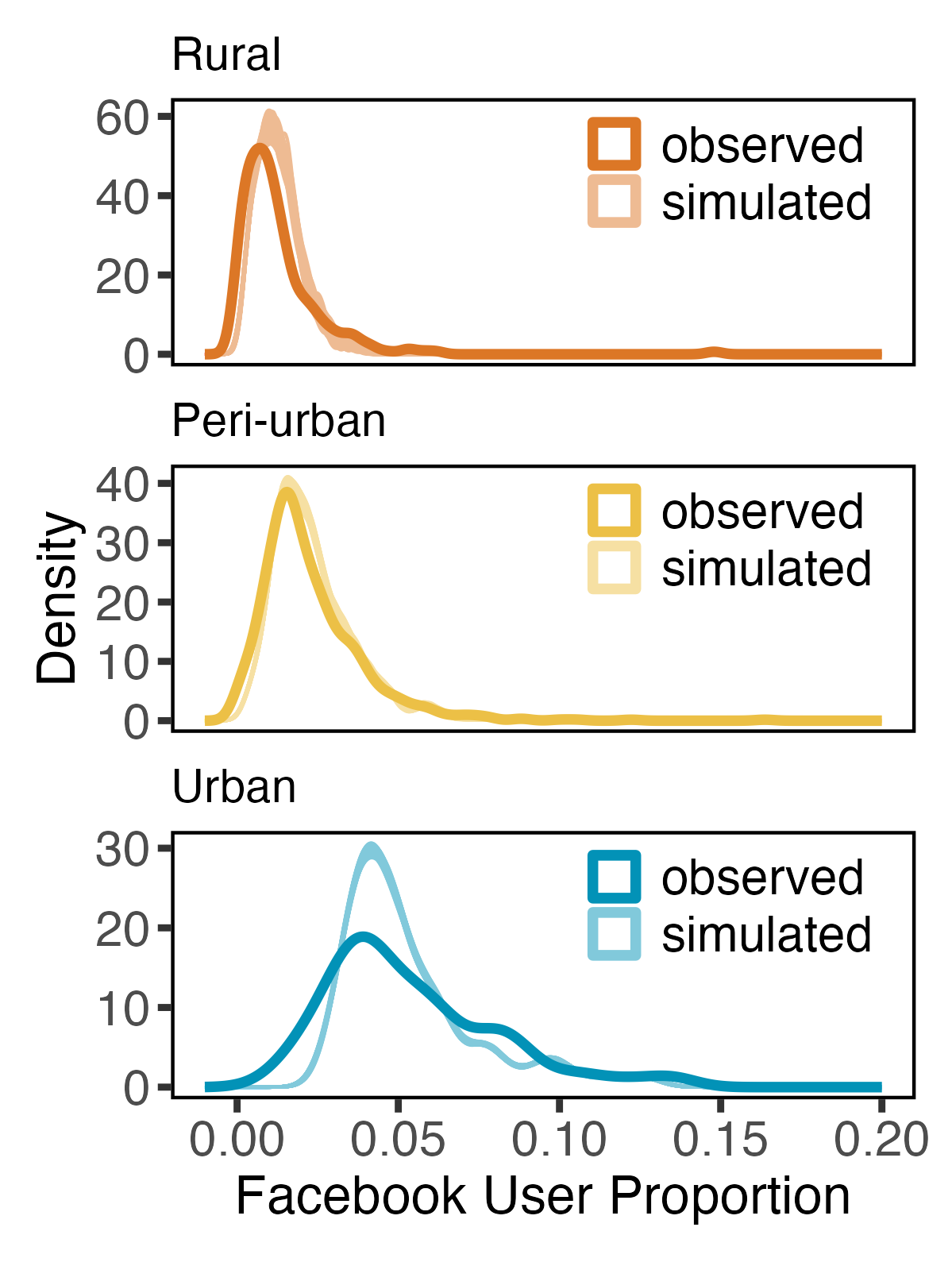}
    }
    \hspace{-0.6cm}
    \subfloat[]{
        \label{fig:dens_ovly_no_sp_over}
        \includegraphics[height=0.34\textheight]
        {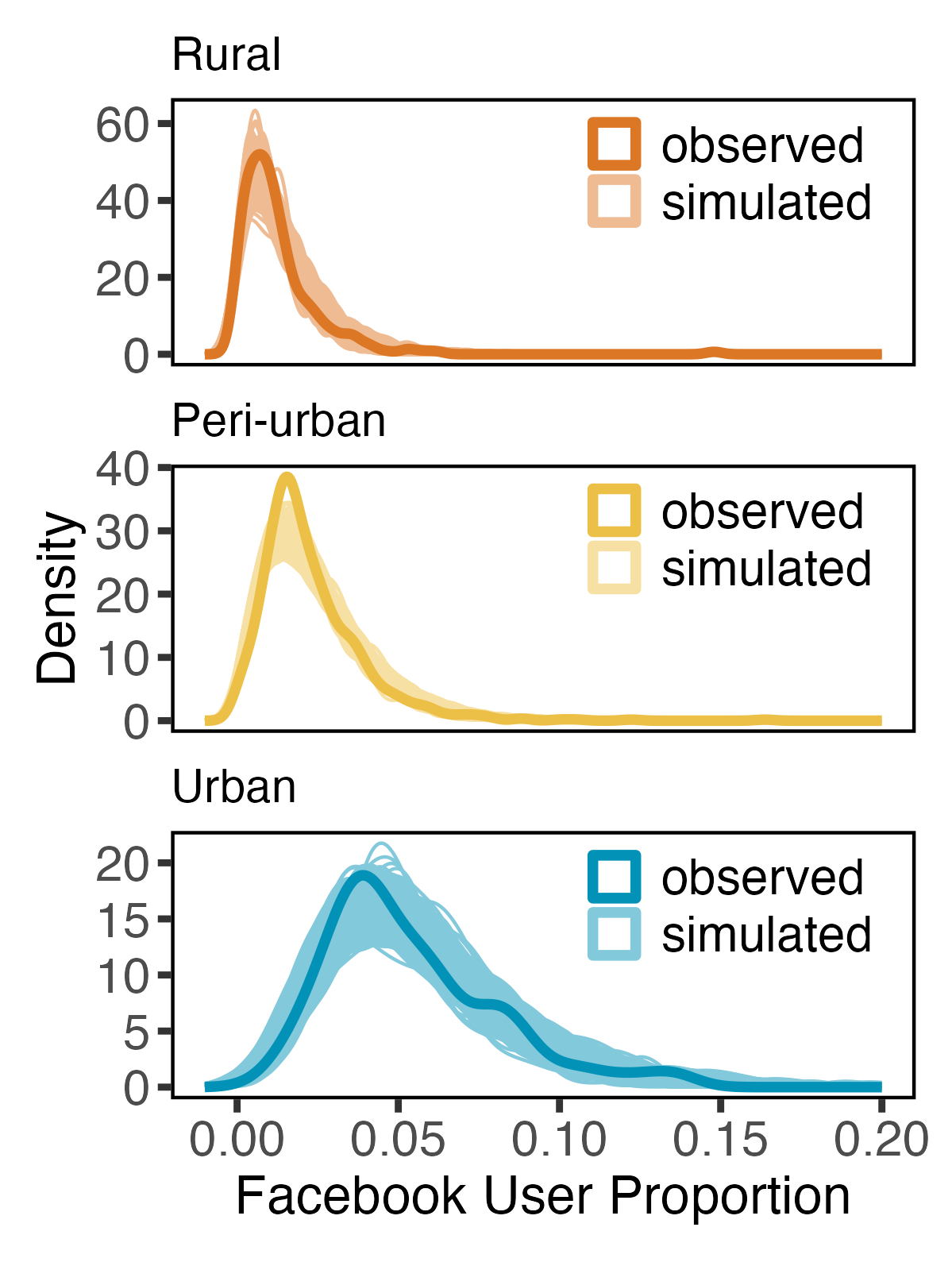}
    }
    \hspace{-0.6cm}
    \subfloat[]{
        \label{fig:dens_ovly_full}
        \includegraphics[height=0.34\textheight]
        {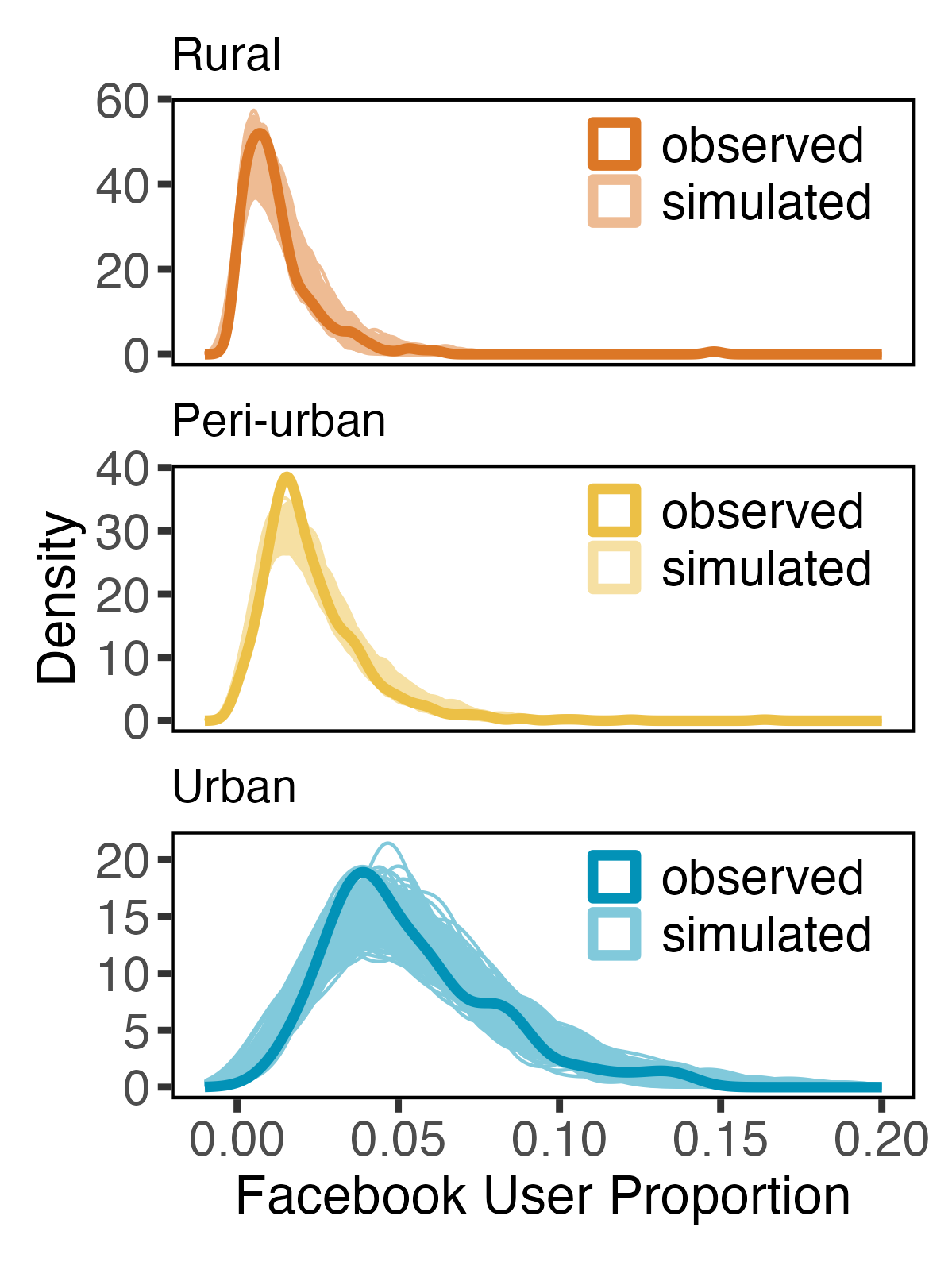}
    }
    \caption{Density plots of Facebook user proportions. The darker lines represent the proportions observed in the train data set, the lighter ones reflect 500 posterior predictive draws from the three models (a) $M_{Bin}$, (b) $M_{BetaBin}$, (c) $M_{full}$.}
    \label{fig:dens_ovly}
\end{figure*}

Figure \ref{fig:traces_no_sp_over} and \ref{fig:pair_plot_no_sp_over} and Table \ref{table:stats_model_betabin} all point to an appropriate MCMC parameter space exploration. 
As expected, the $M_{betabin}$ parameters have significantly larger standard deviations (see Table \ref{table:stats_model_summary}). The means of their posterior samples also markedly change in a few cases, suggesting that the greater model flexibility enables the inference of more accurate point estimates.
Finally, Figure \ref{fig:dens_ovly_no_sp_over} and \ref{fig:psis_diagnostics} confirm that the added overdispersion guarantees that $M_{betabin}$ can better mirror the underlying data distribution. A comparison of the LOO expected logarithm of the pointwise predictive density (LOO-ELPD) for the two models corroborates this further: the model with overdispersion results in a significantly larger ELPD, indicating a better out-of-sample predictive fit\cite{Vehtari2017}.

\subsection{
    \label{sec:model_full}
    Full Model
}

As a last step, we include a spatially explicit component in the model ($M_{full}$), which accounts for correlations between nearby administrative areas that are not already captured by other predictors. A simpler model that included only an intercept term and a spatial correlation component was also tested. This, however, was less accurate than $M_{betabin}$, proving that the predictors by themselves provide important information regarding Facebook usage (these results are available in the Supplementary Code Repository).

To encode the spatial component, we use Hilbert space approximate Gaussian Processes (HSGPs)\cite{Mayol2023}; the significant computational speedup over regular Gaussian Processes (GPs) that this approach offers makes it particularly appealing for scenarios where it is important to rapidly update or test models with new data. However, HSGPs require the selection of two parameters: the number of basis functions $n_{b}$ used to approximate the GPs and the spatial boundaries over which the approximation is valid. A complete analysis of this selection process is beyond the scope of this work, but best practices can be found in \cite{Mayol2023}. Here, we set the approximation boundaries to $\pm5/2 \max(x)$ and $\pm5/2 \max(y)$, where $x$ and $y$ are the (standardised) latitude and longitude positions of the Bayans' centroids. We select $n_{b}$ empirically by comparing models with a progressively increasing number of basis functions (up to $n_{b}$ = 32) through their accuracy on the train data, their LOO-ELPD, and by examining the Pareto $k$ diagnostics (see \ref{sec:supp_hsgp}).

The formulation of the model changes only in the definition of $\overline{p_{FB, i}}$, which becomes:
\begin{align*}
    \overline{p_{FB, i}} &= logit^{-1}(a[U_{i}] + b_{w}[U_{i}] \cdot W_{i} + b_{l}[U_{i}] \cdot logL_{i} + \phi_{16}(\delta,\sigma)) \\[8pt]
\end{align*}
where $\phi_{16}(\delta,\sigma)$ represents the HSGP spatial component constructed with 16 basis functions. $\delta$ and $\sigma$ refer to the length scale and marginal amplitude of the approximated Mat\'ern$_{3/2}$ kernel. Figures \ref{fig:traces_full} and \ref{fig:pair_plot_full} show once again that the parameter space was sampled appropriately. However, the intercept parameters for the three degrees of urbanisation have become significantly correlated. As reported in \ref{table:stats_model_summary}, this results in a higher uncertainty for these variables, while for all other parameters we recover roughly the same standard deviation observed for $M_{betabin}$. Almost all the slope factors have mean values lower than those recovered with the simpler overdispersed model, suggesting that spatial correlations are absorbing a component of the patterns in the data previously explained by the proportion of working age population or nighttime luminosity. In addition, all the $\rho$ parameters decrease, implying a more contained overdispersion effect, a consequence of part of the variability being captured by the HSGP component. This change is reflected in smaller standard deviations for the posterior predictive distribution of the rate of Facebook users for each administrative unit. In particular, the median standard deviation across all Bayans decreases by $\sim11\%$ with respect to its value for $M_{betabin}$, indicating that $M_{full}$ provides predictions with lower uncertainty. 

\begin{table*}[!b]
\centering
\begin{tabular}{
   |>{\raggedright\arraybackslash}m{1.8cm}|
   >{\raggedleft\arraybackslash}m{1.5cm}|
   >{\raggedleft\arraybackslash}m{1.5cm}|
   >{\raggedleft\arraybackslash}m{1.5cm}|
   >{\raggedleft\arraybackslash}m{1.5cm}|
   >{\raggedleft\arraybackslash}m{1.5cm}|
   >{\raggedleft\arraybackslash}m{1.5cm}|
 }
 \hline
 \multicolumn{1}{|>{\centering\arraybackslash}m{1.8cm}|}{\multirow{2}{*}{\textbf{Variable}}} 
 & \multicolumn{2}{>{\centering\arraybackslash}m{3cm}|}{\textbf{$M_{bin}$}}
 & \multicolumn{2}{>{\centering\arraybackslash}m{3cm}|}{\textbf{$M_{betabin}$}}
 & \multicolumn{2}{>{\centering\arraybackslash}m{3cm}|}{\textbf{$M_{full}$}} \\
 \cline{2-7}
 & \multicolumn{1}{>{\centering\arraybackslash}m{1.5cm}|}{\textbf{Mean}} 
 & \multicolumn{1}{>{\centering\arraybackslash}m{1.5cm}|}{\textbf{SD}}
 & \multicolumn{1}{>{\centering\arraybackslash}m{1.5cm}|}{\textbf{Mean}} 
 & \multicolumn{1}{>{\centering\arraybackslash}m{1.5cm}|}{\textbf{SD}}
 & \multicolumn{1}{>{\centering\arraybackslash}m{1.5cm}|}{\textbf{Mean}} 
 & \multicolumn{1}{>{\centering\arraybackslash}m{1.5cm}|}{\textbf{SD}} \\
 \hline
 $a$[1] & -3.832 & 0.007 & -3.950 & 0.105 & -4.123 & 0.189 \\
 $a$[2] & -3.770 & 0.001 & -3.757 & 0.017 & -4.054 & 0.170 \\
 $a$[3] & -3.369 & 0.001 & -3.352 & 0.047 & -3.694 & 0.177 \\
 $b_{w}$[1] & 0.523 & 0.004 & 0.423 & 0.041 & 0.306 & 0.042\\
 $b_{w}$[2] & 0.523 & 0.002 & 0.478 & 0.021 & 0.376 & 0.027 \\
 $b_{w}$[3] & 0.008 & 0.001 & 0.213 & 0.053 & 0.125 & 0.054 \\
 $b_{l}$[1] & 0.533 & 0.010 & 0.376 & 0.145 & 0.511 & 0.129 \\
 $b_{l}$[2] & 0.396 & 0.002 & 0.381 & 0.027 & 0.286 & 0.030\\
 $b_{l}$[3] & 0.263 & 0.001 & 0.160 & 0.026 & 0.126 & 0.026\\
 $\rho$[1] & - & - & 0.007 & 0.001 & 0.006 & 0.001 \\
 $\rho$[2] & - & - & 0.005 & 0.000 & 0.004 & 0.000 \\
 $\rho$[3] & - & - & 0.008 & 0.001 & 0.007 & 0.001 \\
 $\sigma$  & - & - & - & - & 0.352 & 0.094 \\
 $\delta$  & - & - & - & - & 0.839 & 0.294 \\
 \hline
\end{tabular}
\caption{Model parameters inferred from each model. All predictors (including the locations' coordinates) are standardised and the parameters are in standard deviation units. Parameters indices refer to the DUC level with 1 = Rural, 2 = Peri-urban, 3 = Urban.}
\label{table:stats_model_summary}
\end{table*}

\subsection{
    \label{sec:model_comparison}
    Results
}

We use three accuracy metrics evaluated on the test dataset to evaluate the models' predictive capabilities. As each metric weighs errors differently, comparing them allows us to better highlight the models' strength and weaknesses. For each administrative unit we calculate a sample-based version of the absolute error of the median (AEMed) and of the squared error of the mean (SEMean). The former is the absolute difference between the observed test value and the median of the posterior predictive draws. SEMean is instead the squared difference between the observed test value and the mean of the posterior predictive draws. Finally, we use the Continuous Ranked Probability Score (CRPS), a generalisation of mean absolute error that better captures the errors of probabilistic forecasts by taking advantage of information from the whole posterior predictive distributions\cite{Gneiting2007, Jordan2019}. All three metrics are then averaged separately over areas with the same DUC, and the square root of the mean SEMean is considered to map all values to the output's scale. Table \ref{table:metrics_comparison} summarises the results.

\begin{table*}[!h]
\centering
\begin{tabular}{
   |>{\centering\arraybackslash}m{1.8cm}|
   >{\raggedleft\arraybackslash}m{2cm}|
   >{\raggedleft\arraybackslash}m{2cm}|
   >{\raggedleft\arraybackslash}m{2.2cm}|
   >{\raggedleft\arraybackslash}m{1.3cm}|
   >{\raggedleft\arraybackslash}m{1.3cm}|
   >{\raggedleft\arraybackslash}m{1.3cm}|
 }
 \hline
 \multicolumn{1}{|>{\centering\arraybackslash}m{1.8cm}|}{\textbf{Deg. Urb.}}
 & \multicolumn{1}{|>{\centering\arraybackslash}m{2cm}|}{\textbf{Mean Pop.}}
 & \multicolumn{1}{|>{\centering\arraybackslash}m{2cm}|}{\textbf{Mean FB rate}}
 & \multicolumn{1}{|>{\centering\arraybackslash}m{2.2cm}|}{\textbf{Metric}}
 & \multicolumn{1}{>{\centering\arraybackslash}m{1.3cm}|}{\textbf{$M_{bin}$}}
 & \multicolumn{1}{>{\centering\arraybackslash}m{1.3cm}|}{\textbf{$M_{betabin}$}}
 & \multicolumn{1}{>{\centering\arraybackslash}m{1.3cm}|}{\textbf{$M_{full}$}} \\
 \hline
 \multirow{3}{*}{\textbf{Rural}}
 & \multirow{3}{*}{26658}
 & \multirow{3}{*}{0.0115}
 & AEMed (\%) & 44.57 & 41.48 & 38.74 \\
 & & & SEMean (\%) & 54.30 & 54.51 & 51.52 \\
 & & & CRPS (\%) & 40.23 & 29.52 & 27.23 \\
 \hline
 \multirow{3}{*}{\textbf{Peri-urban}}
 & \multirow{3}{*}{42748}
 & \multirow{3}{*}{0.0223}
 & AEMed (\%) & 35.55 & 34.64 & 29.79 \\
 & & & SEMean (\%) & 54.37 & 53.21 & 46.36 \\
 & & & CRPS (\%) & 33.65 & 25.49 & 22.01 \\
 \hline
 \multirow{3}{*}{\textbf{Urban}}
 & \multirow{3}{*}{203762}
 & \multirow{3}{*}{0.0486}
 & AEMed (\%) & 30.70 & 25.49 & 24.16 \\
 & & & SEMean (\%) & 43.03 & 37.46 & 36.34 \\
 & & & CRPS (\%) & 29.94 & 19.18 & 18.25 \\
 \hline
\end{tabular}
\caption{Metrics for each model and DUC. The mean population and Facebook user proportion are shown as a reference. The metrics are expressed in percentage difference from the true Facebook user proportion.}
\label{table:metrics_comparison}
\end{table*}

\begin{figure*}
    \centering
    \includegraphics[height=0.43\textheight]
    {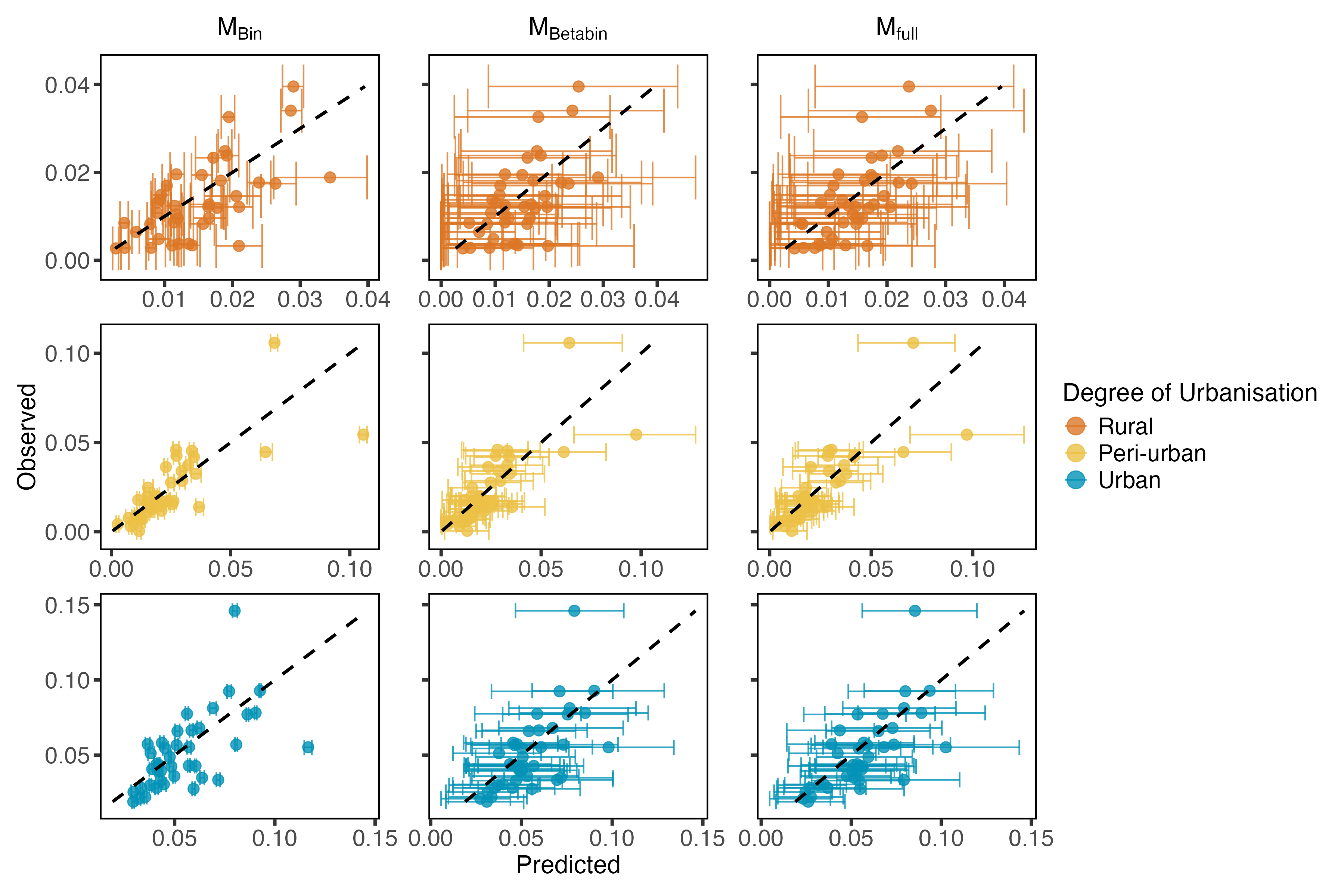}
    \caption{Scatter plots of the observed and predicted Facebook user rates for a subset of administrative area (to improve visibility). Panels are separated by  DUC along the rows and by model along the columns. The error bars identify the 87\% highest density intervals of the posterior predictive distributions.}
    \label{fig:y_rep_y}
\end{figure*}

Figure \ref{fig:y_rep_y} visualizes the model accuracy by providing a comparison between the predicted Facebook users rate (mean and 87\% highest density interval) and the observed values for a subset of administrative unit in the test set. This view clearly highlights the shortcomings of the simple binomial model, which consistently underestimates the prediction uncertainties. Both models that account for overdispersion produce better confidence ranges, which, particularly in the case of rural Bayans, constitute a significant proportion of the true value. This follows from the very small user rate in many of these areas and explains the substantial estimated percentage errors. The figure also reveals that $M_{full}$ performs well for most of the peri-urban and urban administrative units and that the observed errors, particularly in SEMean, is largely driven by a few outliers with unusually large or small observed user rates. In Figure \ref{fig:outliers}, we highlight how these outliers are either peri-urban areas occurring immediately next to an urban area, or small urban areas nestled between several other high density Bayans. In these cases, the anomalous rates likely derive from the imprecise allocation of users between neighboring areas when the signal from the Facebook tiles is distributed to the overlapping regions. This suggests that an automatic user allocation to governmental geographical boundaries rather than to tiles could result in a significant improvement of the model results.

\section{
    \label{sec:discussion}
    Discussion
}

The analysis presented in this work offers several insights into questions of digital access and tries to establish if real-time social media data can be robustly used to estimate population numbers in the event of humanitarian emergencies.

The non-trivial relationship observed in Figure \ref{fig:pair_plot} between population density and the proportion of Facebook users highlights the complex digital access patterns and demographic differences between urban and rural areas. Overall, Facebook usage rapidly rises with increasing population density before flattening around 10\%, with only a few Bayans exceeding this level (for some of which the high values might result from user misallocation as explained above). This behaviour likely reflects a process of rapid technology adoption as tools become accessible even in towns of moderate size. The observed saturation could then be ascribed to an intrinsic maximum acceptance threshold for location sharing among Facebook users (the Philippines estimated Facebook usage\cite{DataReportal2025} is significantly higher than what is observed in the data). When we separate the Bayans by DUC, the picture becomes more subtle as rural areas oppose this trend, with higher population density locations linked to a lower density of users. As mentioned above, this could reflect the socio-economic distinction between more affluent communities living in large estates and denser, poorer realities where digital access is greatly hindered. 

Valuable insights can also be obtained from a closer analysis of the model parameters. Table \ref{table:stats_model_summary} shows that all predictors are robustly associated with the rate of Facebook users, as expected. Higher levels of urbanisation, larger proportions of people of working age, and stronger nighttime luminosities are consistently mapped to a higher proportion of users. Interestingly, this effect is strongest for rural and peri-urban areas, which see the largest shift in user proportion, particularly in response to changes in the nighttime luminosity. The weaker association in more urban areas likely reflects the diminishing effect of electricity access and wealth on the adoption of new technologies at the higher values of luminosity observed for these locations.

Additionally, Figure \ref{fig:pair_plot_full} shows that only in urban areas the contribution of the proportion of working age population and that of the nighttime luminosity are significantly correlated, while they appear to provide orthogonal information for rural areas. This likely encodes the existence of different socio-economic patterns characterised by primary sector occupations less interlinked with the electrical grid.

Finally, the forecast metrics of our models hint at the complexity of the dynamics at play. Including more socio-economic and demographic variables is likely to improve the prediction performance. However, the real-time nature of the population estimates greatly restricts the type of usable information. If interpretability of the results is not required, different modelling approaches, such as tree-based machine learning algorithms or neural network architectures could be used in the attempt to boost the models' forecasting capabilities.

\section{
    \label{sec:limitations}
    Limitations and Strengths
}

Our proposed approach has several limitations resulting from the assumptions introduced in the analysis and to technical challenges in the operationalisation and generalisation of the results. In order to have an appropriate comparison between the users' figures and the true underlying population, we use data from the period of the COVID-19 epidemic. This choice ensures that the users' location at night likely maps onto their place of residence, but, for the results to be sensible, it requires that attitudes towards technology adoption are not sharply affected by the emergency. More generally, we assume that Facebook usage does not follow trends that are not captured by the predictors used in our models. The user count data observed throughout 2021 (see Fig. \ref{fig:suppFB_2021}), however, already hints towards attitude shifts that are likely cultural in nature. Access to long-term historical data would enable the modelling of Facebook adoption trends and the consequent adjustment of future population estimates.

The stationarity assumption for the Facebook user count distribution is even stronger in the context of the data imputation model, where all weekday night data are aggregated throughout 2020. Progressively re-fitting the model on the most recent data could lessen the impact of this assumption and nudge the imputed values towards more realistic estimates.

We also implicitly assume that daily, weekly, and seasonal population movements do not significantly affect the demographic composition of an area or the association of DUC and nighttime luminosity with the density of users. This assumption is unlikely to be fully satisfied as workers move in and out of a business district and tourists visit popular destinations. This limitation could be addressed using alternative data sources that more directly reflect the ever-changing population flow, such as motorway traffic and public transit usage, and rely less on inhabitants' characteristics. As these datasets become available in low- and middle-income countries, models will become more accurate and reactive.

We consider the census data as the ground truth for the population numbers to simplify our analysis. While the Philippines Statistics Authority does not provide official guidance on the matter, census data can in general be affected by sampling and non-sampling errors, which should be included in the model to obtain more accurate uncertainty estimations. 

The operationalisation of the methods we propose here for guiding humanitarian responses also faces some practical challenges. For one, the implementation requires access to proprietary data, a possible hurdle for many organisations. However, Meta and other companies have in the past publicly released some user location information in times of emergency, which would be sufficient to generate population estimates. The technical integration of these methods within existing disaster response systems will pose some difficulties and require dedicated training for humanitarian practitioners and government agencies. Nonetheless, we expect that the simplicity and interpretability of the models presented here will greatly benefit their practical uptake even by non-technical users.

Lastly, our methods can only be generalised to other countries or regions with a significant social media uptake. Indeed, we showed how very low user rates translate to mostly censored signal and provide a much noisier source of information. Limited access to digital infrastructures and marked differences in cultural attitudes towards social media platforms would likely require other real-time approaches, such as the use of mobile phone call detail record data.  

On the other hand, our method has several advantages over these alternatives. Facebook (or more in general social media) data covers the whole globe, ensuring in principle that techniques based on it can be readily extended to other countries and regions. This contrasts with mobile phone data that commonly requires separate access agreements for each service provider. 

The approaches we propose also compare favorably against other traditional methods for SAPEs based on census data and long-term demographic estimates\cite{Bondarenko2025}. Our models allow for new population projections every time a new user data point is released (8 hours), crucial when the movement of people throughout the day, week and year introduces significant changes from the expected resident population. The estimates also respond to changes in the embedded socio-economic variables and can be quickly fit to new official or estimated population data that can be treated as ground truth.

\section{
    \label{sec:conclusions}
    Conclusions
}

The modelling approach developed in this work provides a robust and interpretable way to generate near real-time population numbers using social media user count data and readily available socio-economic proxy predictors. At the same time, we shed light on the necessary assumptions and technical complexities faced when combining these data sources. These are reflected in the sizable uncertainties associated with the inferred population numbers. Using multiple data points for each administrative area (for instance, from multiple weekdays around the census reference date) might help reduce this uncertainty by providing more information in the fitting process. Further improvements include the modelling of technology adoption trends over time, the combination of data from multiple social media sources to balance platform-specific biases, and the addition of stronger social media usage predictors.

It will be important to validate all approaches using data from other disaster-prone regions with different demographic, socio-economic, and cultural characteristics. It is also critical that safeguards are implemented to ensure that users' privacy is protected and that bad-faith actors cannot misuse these tools for the purpose of inappropriately targeting other people. Finally, in-depth analyses of the models are required to ensure that the algorithms are equitable and that outputs well represent each stratum of the population.

\vspace{50pt}

\label{sec:Data and code availability}
\textbf{Data and code availability}
The tile level Facebook user data for this study were obtained through Meta’s AI for Good program. They are not publicly available but information on requesting access can be found at https://ai.meta.com/ai-for-good/datasets/. All other data used in this article are publicly available, with sources listed in the text. They are also collected for ease of access at https://zenodo.org/records/18376171. All the code used to process the data, fit models, and generate visualisations can be found at https://github.com/pandrich/meta-pop-philippines. This contains steps to simulate Facebook user data and test the models.

\label{sec:Author contributions}
\textbf{Author contributions}
A.J.T., S.L. and P.A. worked on the research conceptualisation and design. S.L., Q.D., Z.C., P.A. and H.J. contributed to Facebook data curation. P.A. built the models and conducted the formal analysis and visualisation. S.F. provided support for the statistical analysis. P.A., S.L. and A.J.T. wrote the manuscript. H.J., Q.D., Z.C. and S.F. contributed to the interpretation of the findings and commented on and revised drafts of the manuscript. All authors read and approved the final manuscript. The corresponding authors (P.A. and S.L.) had full access to all the data in the study and had final responsibility for the decision to submit the manuscript for publication.

\label{sec:Acknowledgments}
\textbf{Acknowledgments}
The authors would like to acknowledge the AI for Good program at Meta for providing access to the data. This work was supported by UNFPA (Project 7406521, UNFPA\slash USAID: Strengthening the availability, quality and usability of common operational datasets on population statistics), EPSRC (UKRI grant number EP\slash V002910\slash 2), the Horizon Europe (UKRI grant number 10041831), and the National Institute for Health (MIDAS Mobility R01AI160780). We thank Romesh Silva and Mallika Snyder from UNFPA for their valuable inputs. The funders of the study had no role in the study design, data collection, data analysis, data interpretation, or writing of the report.

\label{sec:Ethics declarations}
\textbf{Ethics declarations}
Ethical approval was obtained from the Research Ethics and Governance Office of the University of Southampton (ERGO II: 87924). The authors declare no conflict of interest.

\clearpage

% \putbib[biblio]   % prints only items cited in this bibunit from refs.bib
% \end{bibunit}

\printbibliography
\end{refsection}

\pagebreak
% \widetext
\begin{center}
\textbf{\large Supplementary Materials}
\end{center}

\setcounter{section}{0}
\setcounter{equation}{0}
\setcounter{figure}{0}
\setcounter{table}{0}
\setcounter{page}{1}
\makeatletter
\renewcommand{\thesection}{S.\arabic{section}}
\renewcommand{\theequation}{S.\arabic{equation}}
\renewcommand{\thefigure}{S.\arabic{figure}}
\renewcommand{\thetable}{S.\MakeUppercase{\roman{table}}}
% \renewcommand{\bibnumfmt}[1]{[S#1]}
% \renewcommand{\citenumfont}[1]{S#1}

% \begin{bibunit}[naturemag]
\begin{refsection}

\section{
    \label{sec:supp_fb}
    Additional information on Facebook data
}

\begin{figure*}[!h]
    \centering
    \subfloat[]
    {
        \label{fig:suppFB_2021}
        \includegraphics[height=0.4\textheight]
        {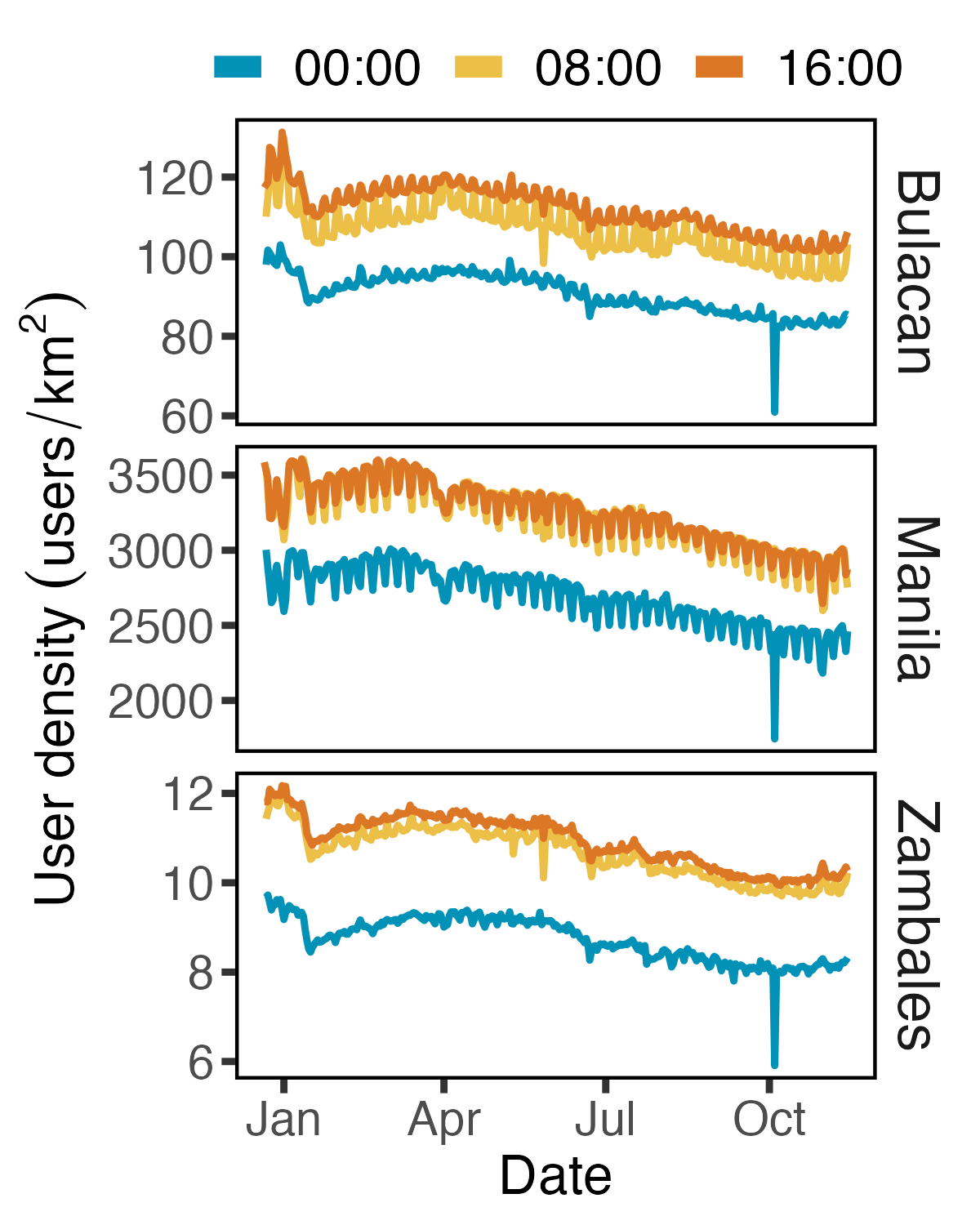}
    }
    \hspace{-0.7cm}
    \subfloat[]{
        \label{fig:supp_FB_miss_after_imputation}
        \includegraphics[height=0.4\textheight]
        {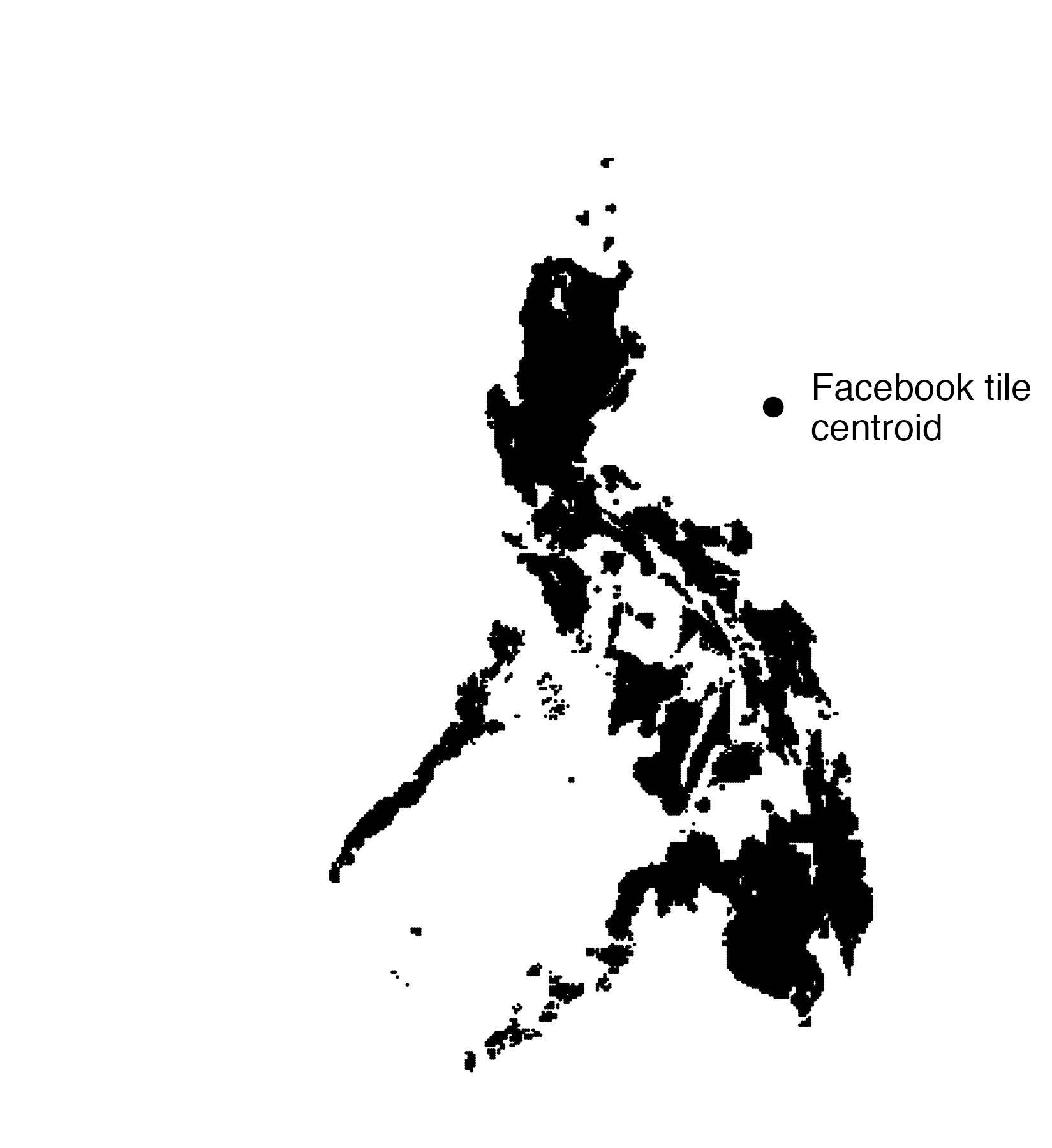}
    }
    \caption{(a) Facebook user density in the same municipalities shown in Figure \ref{fig:fb_over_time} for 2021. (b) Location of the Bing tile centroids showing an almost perfect coverage of the whole country. }
    \label{fig:suppFB}
\end{figure*}

\begin{figure*}[!h]
    \centering
    \includegraphics[height=0.4\textheight]
    {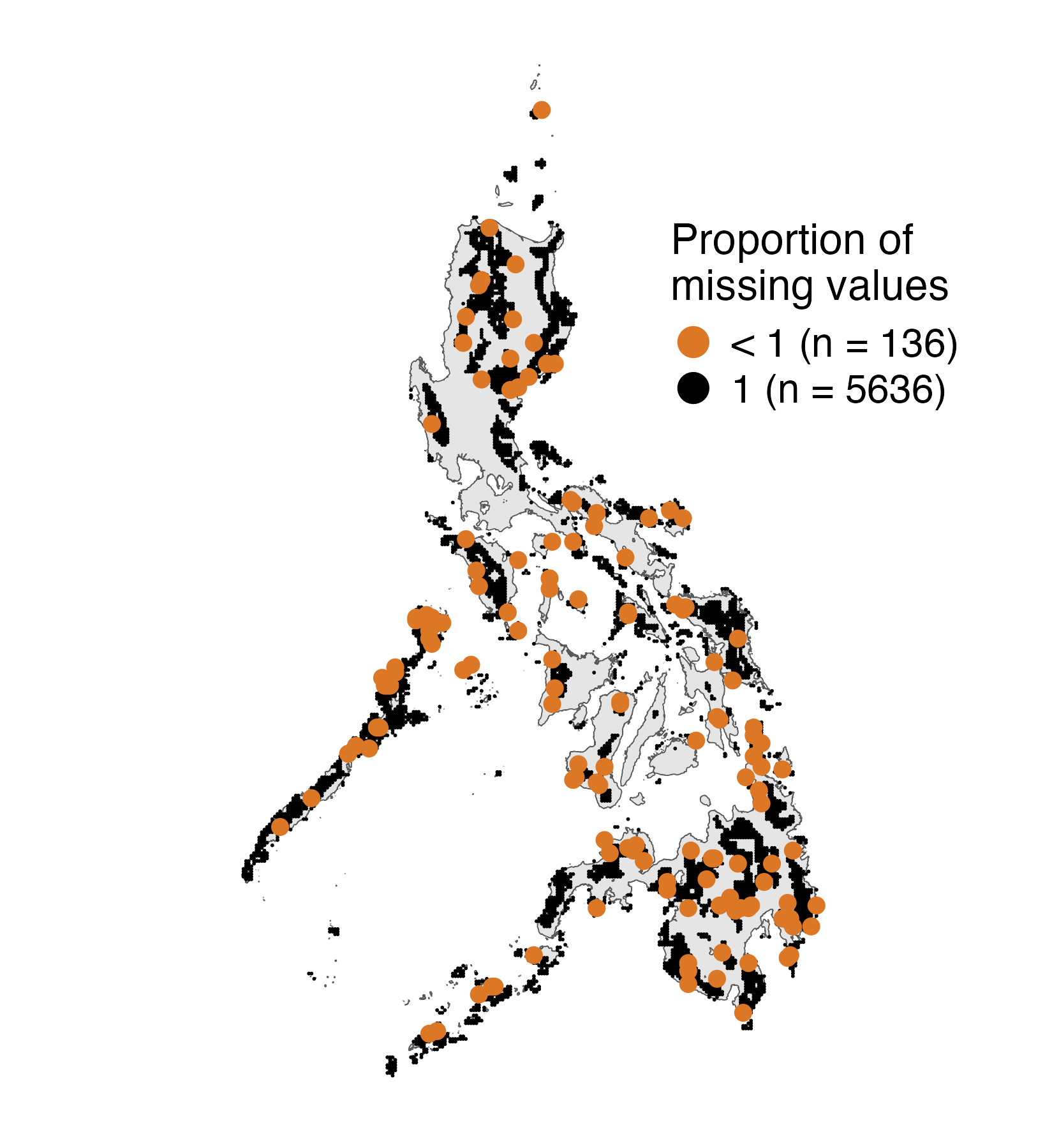}
    \caption{Centroid location of the Bing tiles with missing value in the nighttime data window on 4 May 2020. Areas with no observed data for all of 2020 and those at least one non-missing data entry are marked in black and orange respectively. The legend reports the number of tiles (n) in each category}
    \label{fig:map_missing}
\end{figure*}

\FloatBarrier
\clearpage

\section{
    \label{sec:imputation_diagnostics}
    Imputation Model Diagnostics
}

\begin{figure*}[!h]
    \centering
    \includegraphics[width=1\textwidth]
    {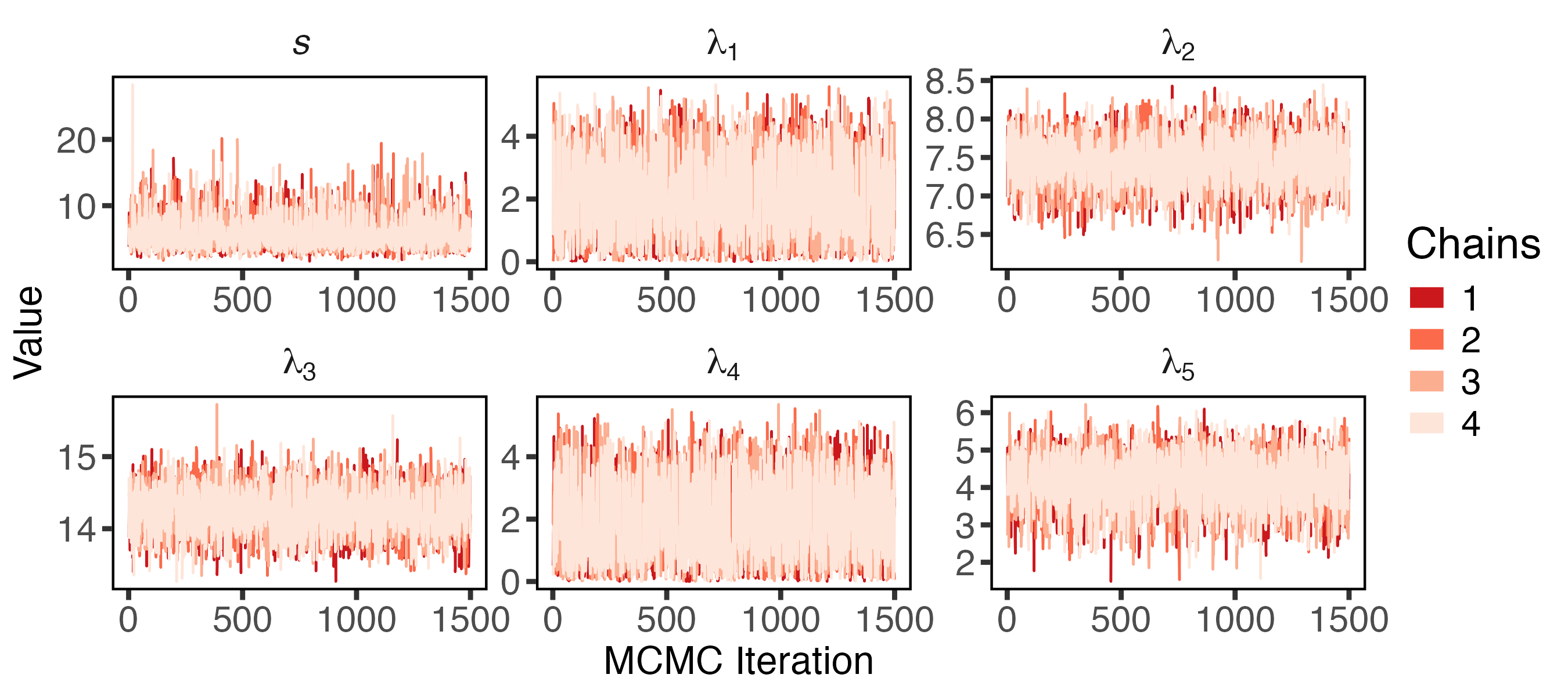}
    \caption{Markov Chain Monte Carlo (MCMC) trace plots of the sampled values for the parameters of the imputation model fitted to data simulated for six tiles (the scale parameter for tile 6 is not presented to simplify the visualization). All parameters show a healthy sampling behaviour.}
    \label{fig:imp_trace_plots_sim}
\end{figure*}

\begin{figure*}[!h]
    \centering
    \includegraphics[width=1\textwidth]
    {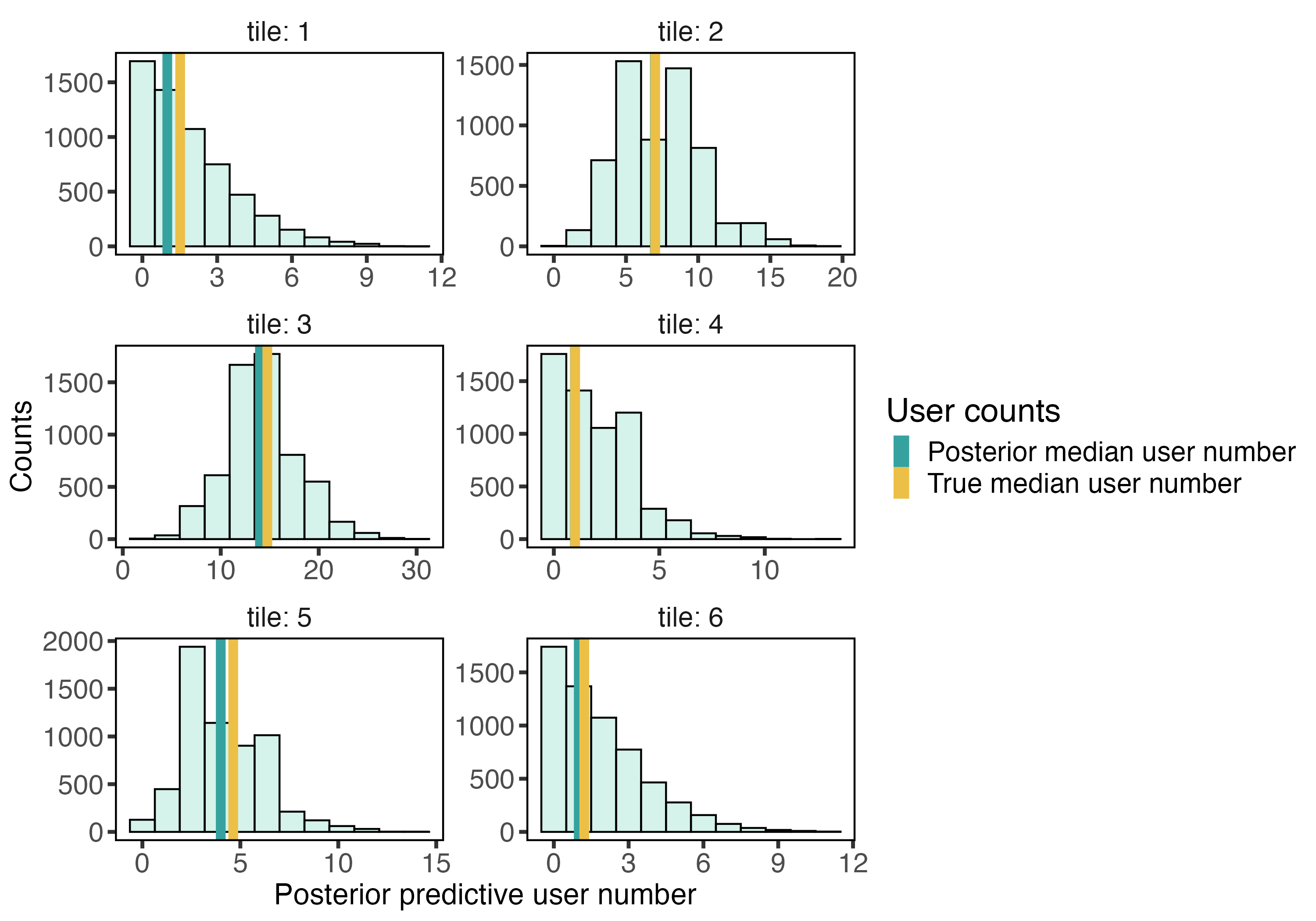}
    \caption{Histograms of user count predictions generated by the imputation model for the six simulated tiles. The vertical lines represent the median user count inferred by the model (green) and the median of the true pre-censoring simulated data (yellow). We see that the model very closely recovers the expected true values.}
    \label{fig:imp_ppc_sim}
\end{figure*}

\begin{figure*}[!h]
    \centering
    \includegraphics[width=1\textwidth]
    {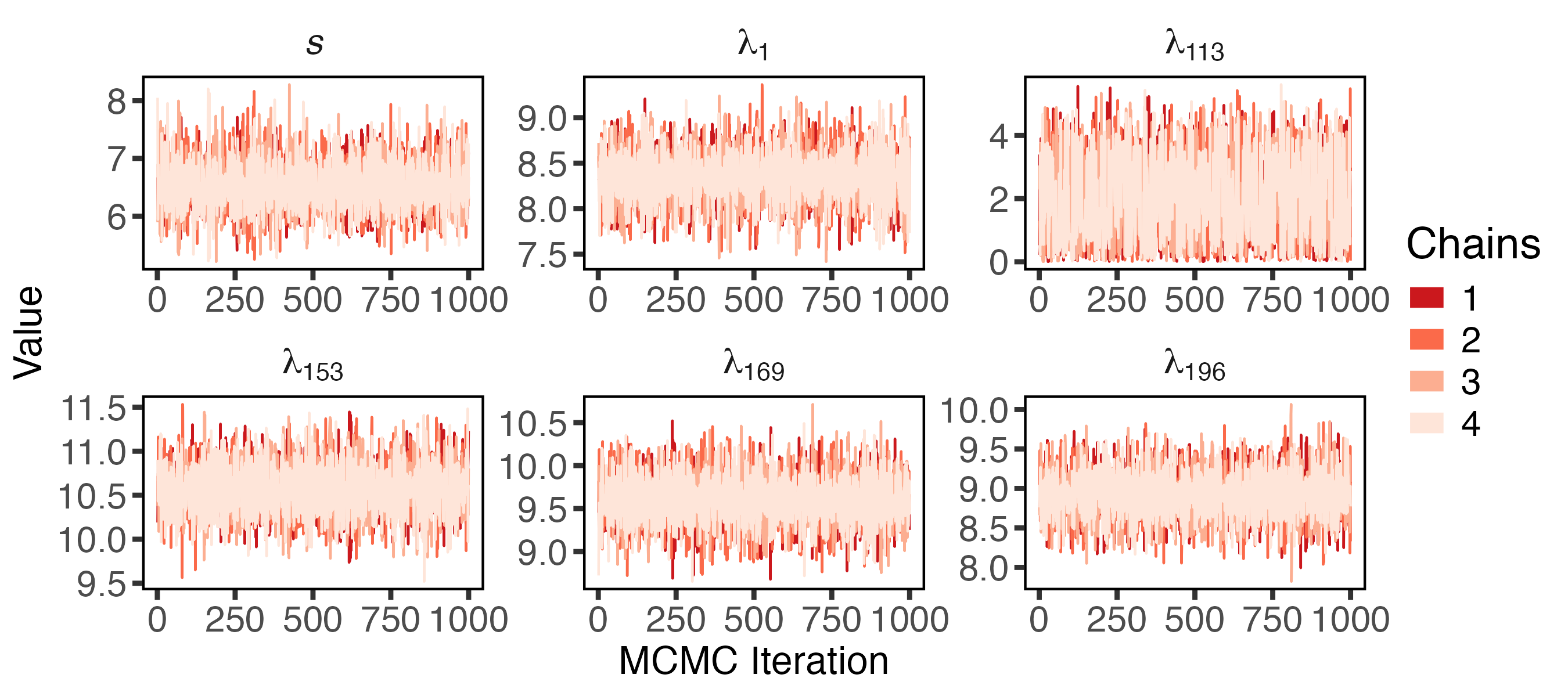}
    \caption{
    MCMC trace plots of the sampled values for the parameters of the imputation model fitted to the real data. Trace plots for parameters of the imputation model fitted to the real data. Shown are the traces for the scale parameter of the hyperprior distribution and for the scale parameter for 5 representative tiles. All parameters show a healthy sampling behaviour.}
    \label{fig:imp_trace_plots}
\end{figure*}

\begin{figure*}[!h]
    \centering
    \includegraphics[width=1\textwidth]
    {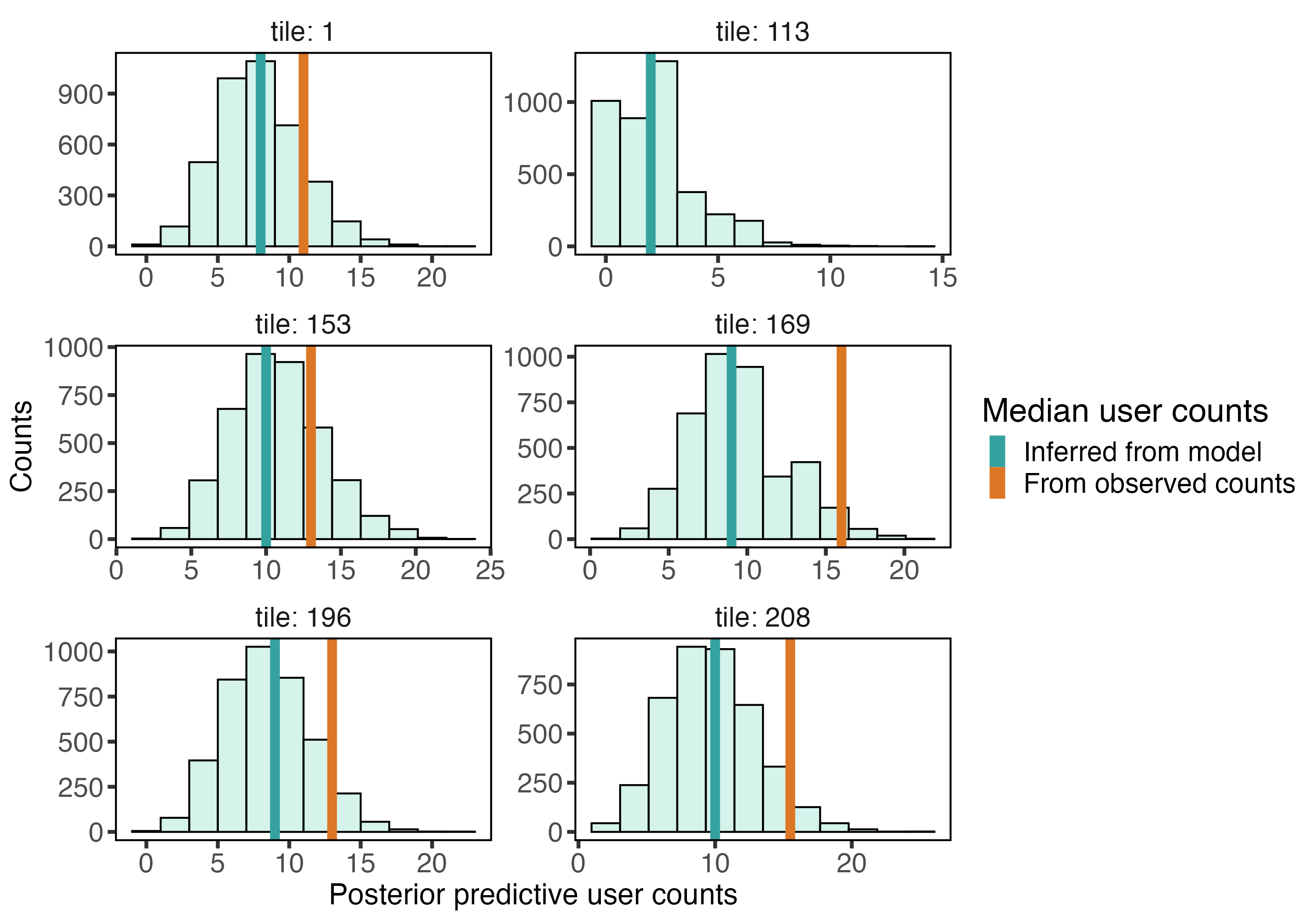}
    \caption{Histograms of user count predictions generated by an imputation model equivalent to that presented in the main text but without hierarchical structure (and consequently no pooling effect between tiles). Here again we see that for all tiles with partially observed data, the model-inferred user counts median (green vertical line) is lower than that of the observed values (orange vertical line). The close similarity of these histograms to those in Figure \ref{fig:imp_ppc} suggests that the information contained in the data for each tile is enough to overwhelm any pooling effect.}
    \label{fig:imp_ppc_non_hier}
\end{figure*}

\FloatBarrier

\section{
    \label{sec:supp:data_mapping}
    Mapping of preprocessed data
}

\begin{figure*}[!h]
    \centering
    \subfloat[]
    {
        \label{fig:var_map_all}
        \includegraphics[height=0.43\textheight]
        {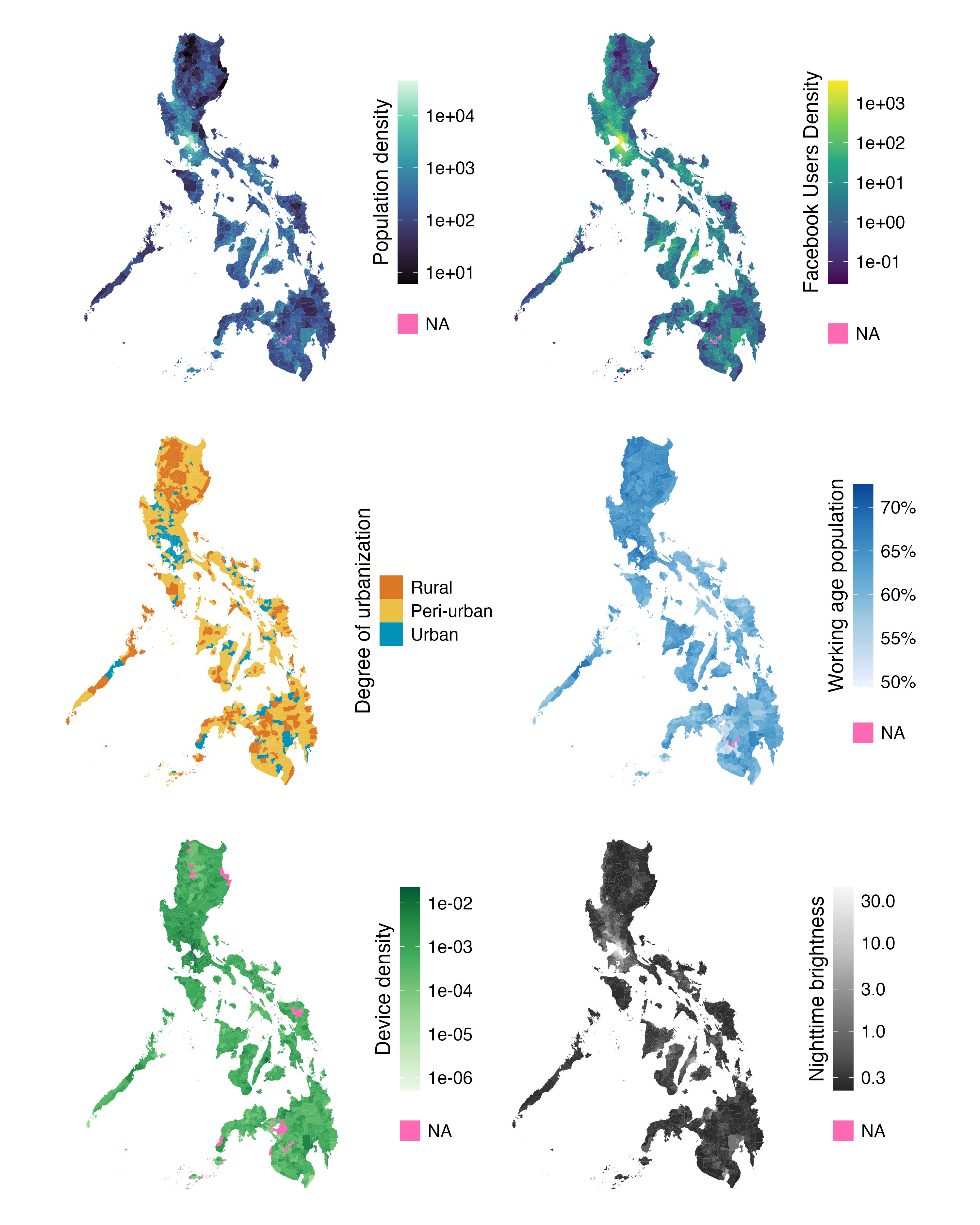}
    }
    \hspace{-0.6cm}
    \subfloat[]{
        \label{fig:var_map_manila}
        \includegraphics[height=0.43\textheight]
        {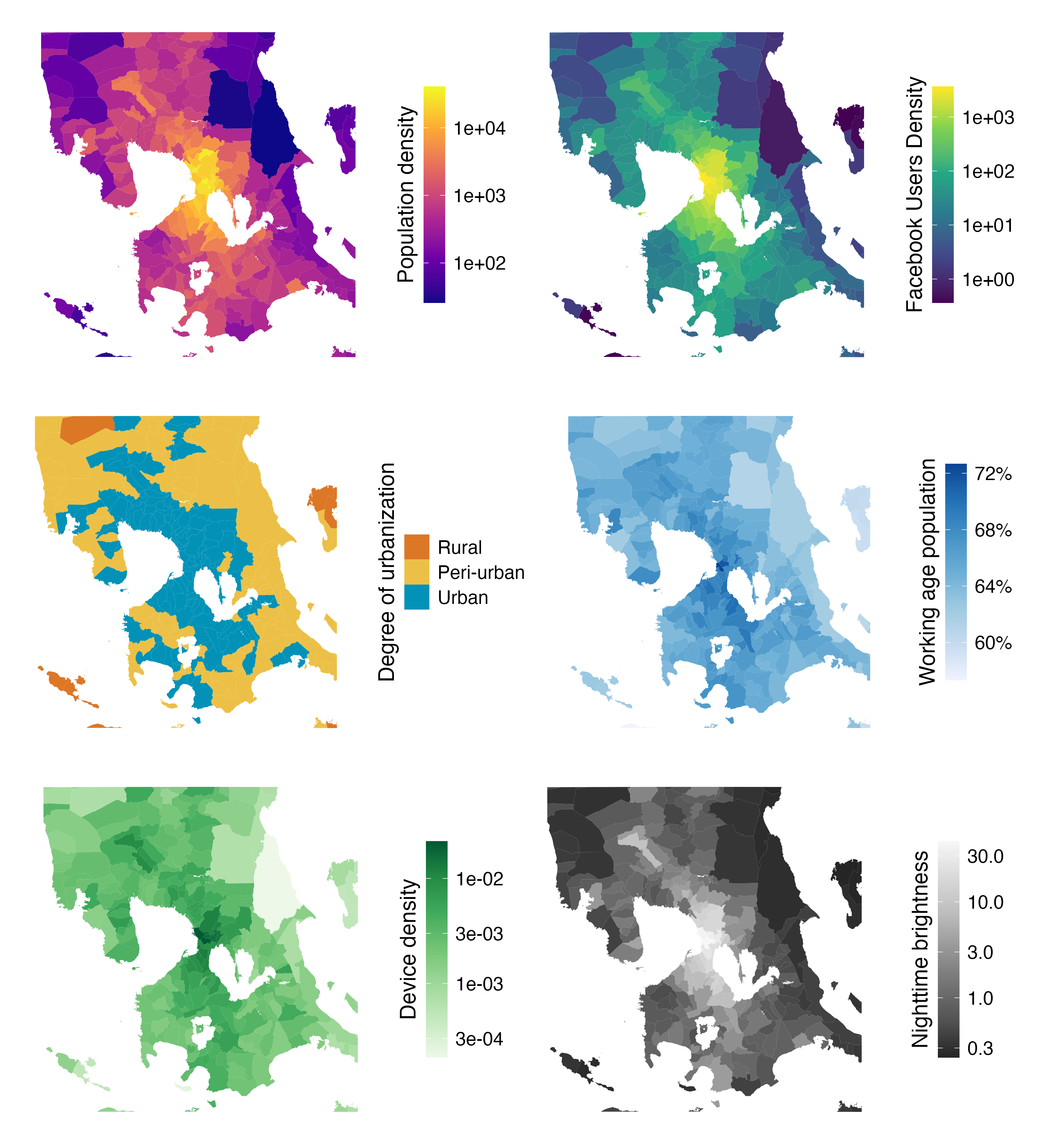}
    }
    \caption{Maps of all the variables used in this work for (a) the whole country and (b) the region surrounding the capital, Manila. In (a), a small area characterised by missing data stands out in the south. Change in the geographic classification of this area between 2020 and 2023 (the reference date of the geospatial dataset used here) prevents matching with the census figures. Missing population data has a direct (on the working age population figures) and indirect (on variables calculated as densities) impact on several indicators. As the Bayan population stock is the reference "ground truth" these areas cannot be included in our analysis. Additional areas lack information for Ookla's network data.}
    \label{fig:var_map}
\end{figure*}

\clearpage
\FloatBarrier

\section{
    \label{sec:supp:model_development}
    Model Development
}

\begin{figure*}[!h]
    \centering
    \includegraphics[width=0.5\textwidth]
    {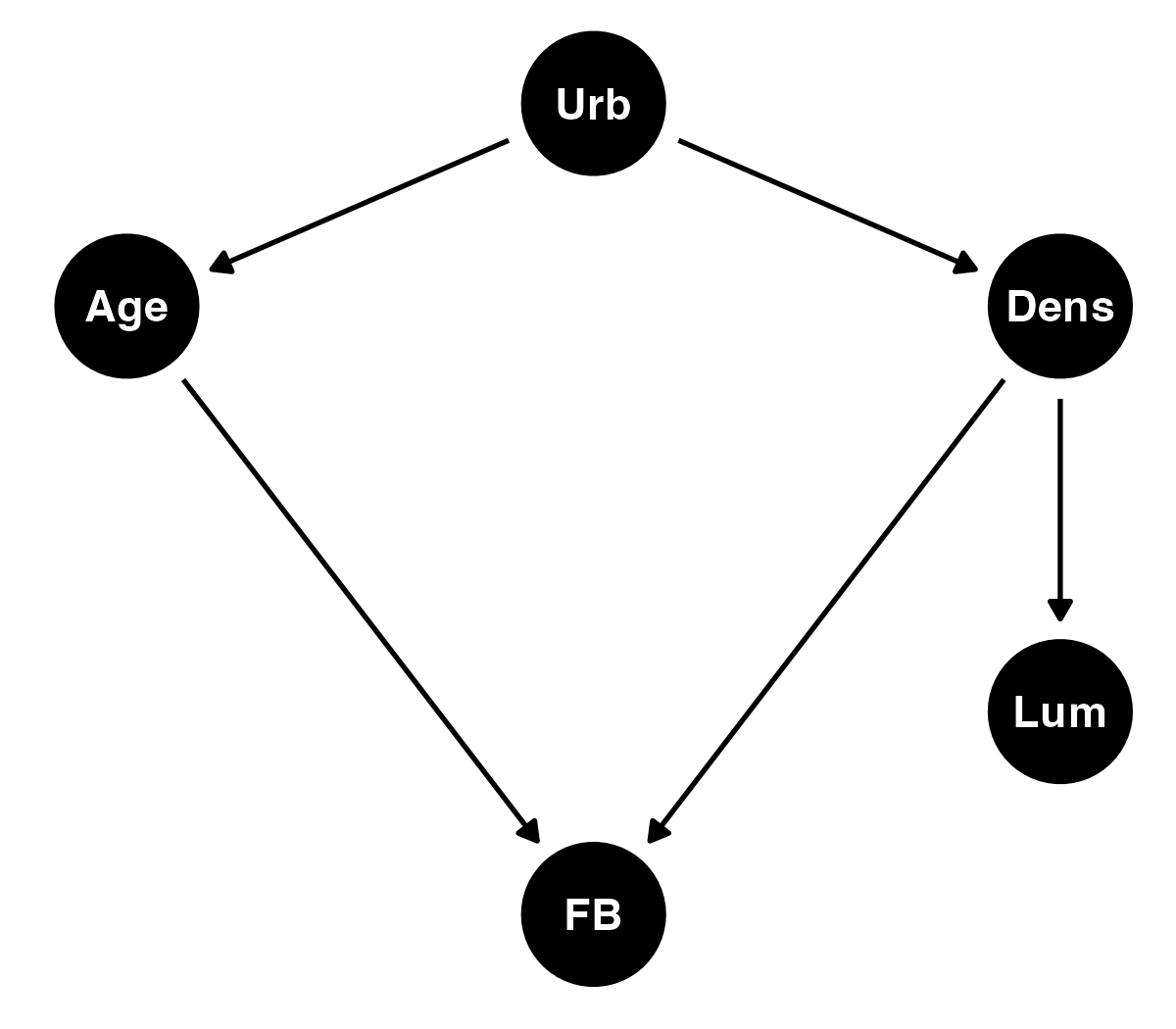}
    \caption{Directed acyclic graph used to develop data simulations and to support the interpretation of the model outputs. The graph nodes are labeled as follow: \texttt{Urb} = Degree of Urbanisation, \texttt{Age} = Proportion of working age people, \texttt{Dens} = Population Density, \texttt{Lum} = Nighttime luminosity, and \texttt{FB} = Proportion of Facebook users.}
    \label{fig:dag}
\end{figure*}

\begin{figure*}[!h]
    \centering
    \includegraphics[width=1\textwidth]
    {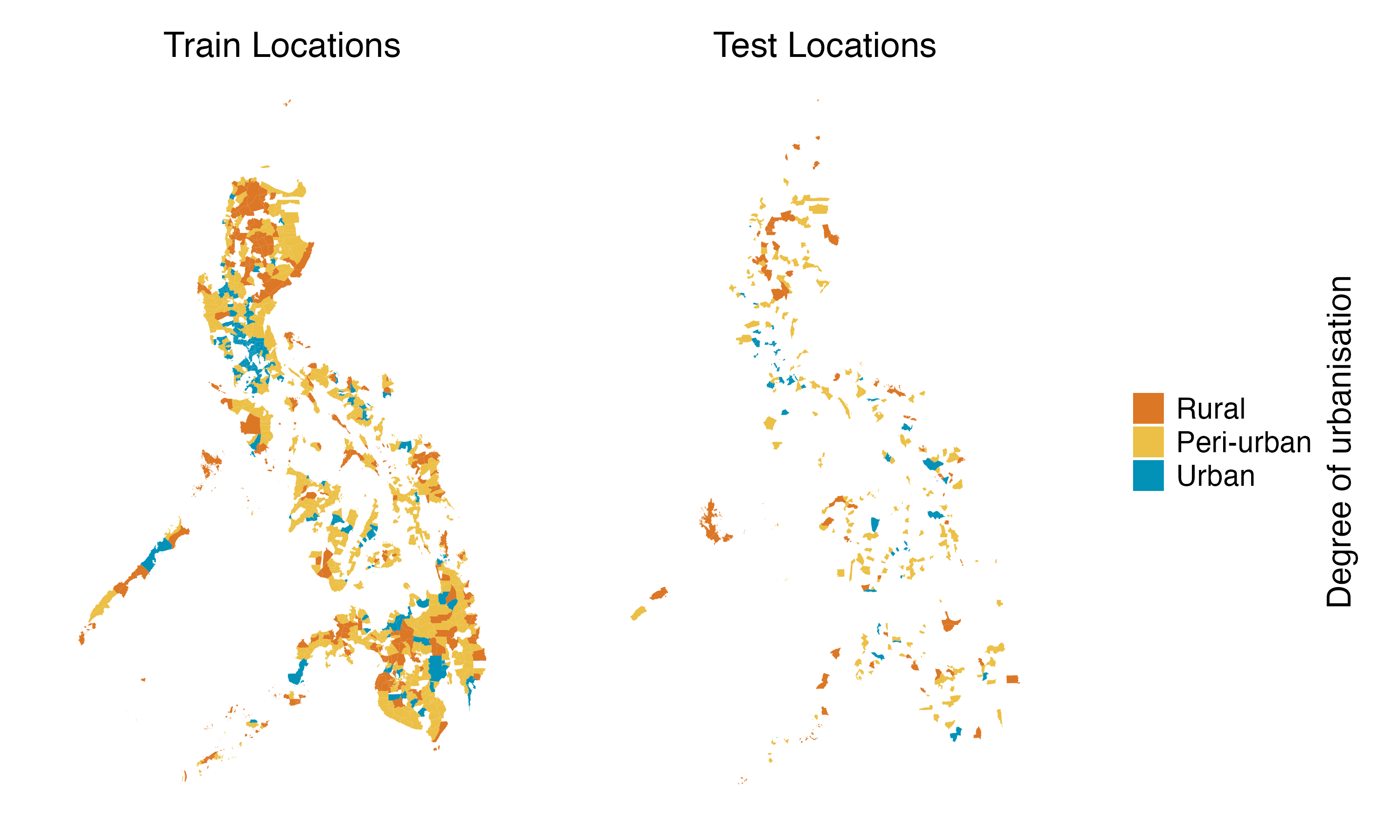}
    \caption{Maps of the municipalities used as model train (left) and test (right) locations.}
    \label{fig:train_test_split}
\end{figure*}

\clearpage
\FloatBarrier

\subsection{
    \label{sec:supp_hsgp}
    HSGP - Basis Selection
}

We use an empirical approach to select the number of basis functions $n_{b}$ for the Hilbert Space Gaussian Process approximation.
To start, we implement a series of models with progressively increasing values of $n_{b}$ (4, 8, 16, 32) to test how this parameter affects the model fits. We limit the search to at most 32 basis functions to maintain the model computationally manageable and because the outputs do not drastically change as $n_{b}$ increases beyond this level.

\begin{figure*}[!h]
    \centering
    \includegraphics[width=1\textwidth]
    {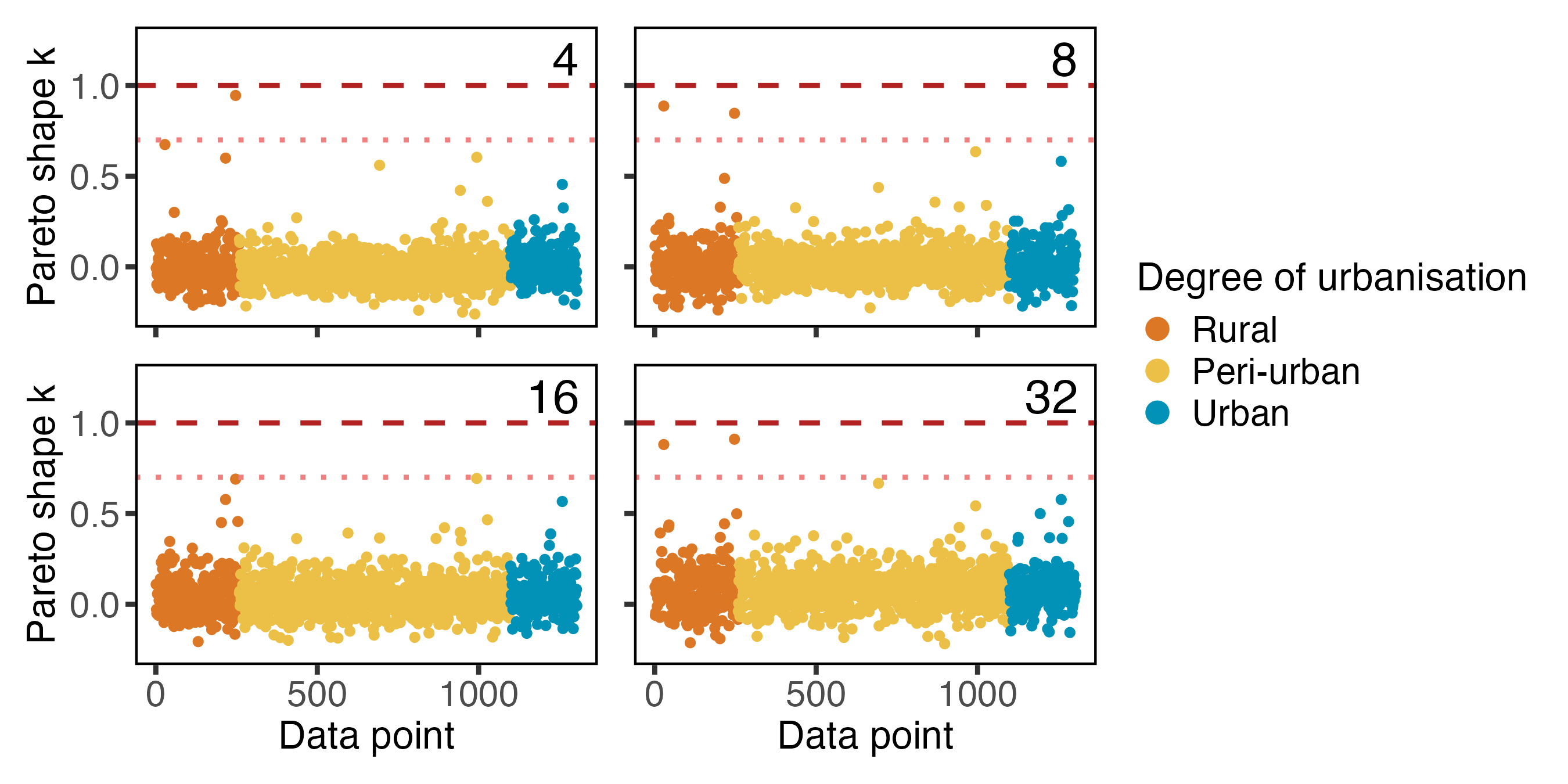}
    \caption{Pareto k diagnostics plots for $M_{full}$ and a progressively increasing number $n_{b}$ of basis functions (as indicated in the top right corner of each plot). Changing $n_{b}$ does not result in large differences but  $n_{b} = 16$ results in the most robust model with all data points showing k values below the 0.7 threshold\cite{Vehtari2017}.}
    \label{fig:psis_diagnostics_hsgp}
\end{figure*}

In Table \ref{table:basis_search}, we show the model metrics calculated using the training dataset. There is no clear pattern, but the model accuracy appears to improve as $n_{b}$ increases for peri-urban and urban areas, but it decreases for rural areas. An investigation of the Pareto $k$ diagnostics provides additional information on the models behaviour (see Fig. \ref{fig:psis_diagnostics_hsgp}). As the number of basis functions increases, the number of points with larger $k$ also grows, possibly indicating a tendency of the models towards overfitting. 

The best model according to the LOO-ELPD metric is the one with 32 basis functions, followed by those with 16, 8, and 4 basis functions that have 10.9, 34.3 and 35.0 lower LOO-ELPD, respectively. From the information we derive from these diagnostics, we select to focus our analysis on the model with 16 basis functions, which combines very light computational requirements with good accuracy. 

We note that, if prediction is the main goal of the model, a more rigorous selection of the optimal $n_{b}$ could be obtained by splitting the data set into three subsets (train, validation, and test) and by selecting the parameter value resulting in the best accuracy metrics calculated on the validation set.

\begin{table*}[!h]
\centering
\begin{tabular}{
   |>{\centering\arraybackslash}m{2.2cm}|
   >{\raggedleft\arraybackslash}m{2.2cm}|
   >{\raggedleft\arraybackslash}m{1.5cm}|
   >{\raggedleft\arraybackslash}m{1.5cm}|
   >{\raggedleft\arraybackslash}m{1.5cm}|
   >{\raggedleft\arraybackslash}m{1.5cm}|
 }
 \hline
 \multicolumn{1}{|>{\centering\arraybackslash}m{2.2cm}|}{\multirow{2}{*}{\textbf{Deg. Urb.}}} 
 & \multicolumn{1}{|>{\centering\arraybackslash}m{2.2cm}|}{\multirow{2}{*}{\textbf{Metric}}} 
 & \multicolumn{4}{|>{\centering\arraybackslash}m{7cm}|}{\textbf{Number of basis functions}} \\
 \cline{3-6}
 & & \multicolumn{1}{|>{\centering\arraybackslash}m{1.5cm}|}{\textbf{4}}
 & \multicolumn{1}{>{\centering\arraybackslash}m{1.5cm}|}{\textbf{8}}
 & \multicolumn{1}{>{\centering\arraybackslash}m{1.5cm}|}{\textbf{16}}
 & \multicolumn{1}{>{\centering\arraybackslash}m{1.5cm}|}{\textbf{32}} \\
 \hline
 \multirow{3}{*}{\textbf{Rural}}
 & AEMed (\%) & 38.63 & 37.90 & 38.75 & 39.46 \\
 & SEMean (\%) & 51.13 & 50.34 & 51.54 & 52.69 \\
 & CRPS (\%) & 27.71 & 27.32 & 27.65 & 28.01 \\
 \hline
 \multirow{3}{*}{\textbf{Peri-urban}}
 & AEMed (\%) & 31.09 & 30.71 & 29.66 & 29.87 \\
 & SEMean (\%) & 47.71 & 47.22 & 46.23 & 45.79\\
 & CRPS (\%) & 22.94 & 22.70 & 21.95 & 21.88\\
 \hline
 \multirow{3}{*}{\textbf{Urban}}
 & AEMed (\%) & 24.21 & 24.43 & 23.85 & 23.44 \\
 & SEMean (\%) & 35.36 & 35.84 & 36.05 & 35.98 \\
 & CRPS (\%) & 17.98 & 18.21 & 18.12 & 17.93 \\
 \hline
\end{tabular}
\caption{Test metrics for $M_{full}$ calculated separately for each degree of urbanisation and for implementations with 4, 8, 16, and 32 basis functions.}
\label{table:basis_search}
\end{table*}

\clearpage
\FloatBarrier

\section{
    \label{sec:supp:model_diagnostics}
    Facebook Uptake Model Diagnostics
}

\begin{table*}[!h]
\centering
\begin{tabular}{
   |>{\raggedright\arraybackslash}m{1.8cm}||
   >{\raggedleft\arraybackslash}m{1.8cm}|
   >{\raggedleft\arraybackslash}m{1.8cm}|
   >{\raggedleft\arraybackslash}m{1.8cm}|
   >{\raggedleft\arraybackslash}m{1.8cm}|
 }
 \hline
 \rowcolor{SeaGreen3!30!}
 \multicolumn{1}{|>{\centering\arraybackslash}m{1.8cm}||}{\textbf{Variable}} 
 & \multicolumn{1}{>{\centering\arraybackslash}m{1.8cm}|}{\textbf{Mean}} 
 & \multicolumn{1}{>{\centering\arraybackslash}m{1.8cm}|}{\textbf{Std. Dev.}}
 & \multicolumn{1}{>{\centering\arraybackslash}m{1.8cm}|}{\textbf{N. Eff.}}
 & \multicolumn{1}{>{\centering\arraybackslash}m{1.8cm}|}{\textbf{Rhat}}\\
 \hline
 $a$[1] & -3.832 & 0.007 & 1261 & 1.001\\
 $a$[2] & -3.770 & 0.001 & 6695 & 0.999 \\
 $a$[3] & -3.369 & 0.001 & 5893 & 1.000 \\
 $b_{w}$[1] & 0.523 & 0.004 & 3288 & 1.000 \\
 $b_{w}$[2] & 0.523 & 0.002 & 5120 & 0.999 \\
 $b_{w}$[3] & 0.008 & 0.002 & 5558 & 1.000 \\
 $b_{l}$[1] & 0.533 & 0.010 & 1274 & 1.002 \\
 $b_{l}$[2] & 0.396 & 0.002 & 5539 & 1.000 \\
 $b_{l}$[3] & 0.263 & 0.001 & 6339 & 1.000 \\
 \hline\hline
\end{tabular}
\caption{Summary table for the main parameters of the $M_{bin}$ model. The effective sample size (N. Eff.) and Rhat statistics very close to 1.0 are indications of a good sampling behaviour.}
\label{table:stats_model_bin}
\end{table*}

\begin{table*}[!h]
\centering
\begin{tabular}{
   |>{\raggedright\arraybackslash}m{1.8cm}||
   >{\raggedleft\arraybackslash}m{1.8cm}|
   >{\raggedleft\arraybackslash}m{1.8cm}|
   >{\raggedleft\arraybackslash}m{1.8cm}|
   >{\raggedleft\arraybackslash}m{1.8cm}|
 }
 \hline
 \rowcolor{SeaGreen3!30!}
 \multicolumn{1}{|>{\centering\arraybackslash}m{1.8cm}||}{\textbf{Variable}} 
 & \multicolumn{1}{>{\centering\arraybackslash}m{1.8cm}|}{\textbf{Mean}} 
 & \multicolumn{1}{>{\centering\arraybackslash}m{1.8cm}|}{\textbf{Std. Dev.}}
 & \multicolumn{1}{>{\centering\arraybackslash}m{1.8cm}|}{\textbf{N. Eff.}}
 & \multicolumn{1}{>{\centering\arraybackslash}m{1.8cm}|}{\textbf{Rhat}}\\
 \hline
 $a$[1] & -3.950 & 0.105 & 2538 & 1.001\\
 $a$[2] & -3.757 & 0.017 & 4738 & 1.000 \\
 $a$[3] & -3.352 & 0.047 & 4856 & 0.999 \\
 $b_{w}$[1] & 0.423 & 0.041 & 5837 & 1.000 \\
 $b_{w}$[2] & 0.478 & 0.021 & 4372 & 1.000 \\
 $b_{w}$[3] & 0.213 & 0.053 & 2933 & 1.001 \\
 $b_{l}$[1] & 0.376 & 0.145 & 2482 & 1.002 \\
 $b_{l}$[2] & 0.381 & 0.027 & 4905 & 1.000 \\
 $b_{l}$[3] & 0.160 & 0.026 & 3265 & 1.001 \\
 $\rho$[1] & 0.007 & 0.001 & 5350 & 1.000 \\
 $\rho$[2] & 0.005 & 0.000 & 5082 & 1.000 \\
 $\rho$[3] & 0.008 & 0.001 & 5225 & 0.999 \\
 \hline\hline
\end{tabular}
\caption{Summary table for the main parameters of the $M_{betabin}$ model. The effective sample size (N. Eff.) and Rhat statistics very close to 1.0 are indications of a good sampling behaviour.}
\label{table:stats_model_betabin}
\end{table*}

\begin{table*}[!h]
\centering
\begin{tabular}{
   |>{\raggedright\arraybackslash}m{1.8cm}||
   >{\raggedleft\arraybackslash}m{1.8cm}|
   >{\raggedleft\arraybackslash}m{1.8cm}|
   >{\raggedleft\arraybackslash}m{1.8cm}|
   >{\raggedleft\arraybackslash}m{1.8cm}|
 }
 \hline
 \rowcolor{SeaGreen3!30!}
 \multicolumn{1}{|>{\centering\arraybackslash}m{1.8cm}||}{\textbf{Variable}} 
 & \multicolumn{1}{>{\centering\arraybackslash}m{1.8cm}|}{\textbf{Mean}} 
 & \multicolumn{1}{>{\centering\arraybackslash}m{1.8cm}|}{\textbf{Std. Dev.}}
 & \multicolumn{1}{>{\centering\arraybackslash}m{1.8cm}|}{\textbf{N. Eff.}}
 & \multicolumn{1}{>{\centering\arraybackslash}m{1.8cm}|}{\textbf{Rhat}}\\
 \hline
 $a$[1] & -4.123 & 0.189 & 1478 & 1.000\\
 $a$[2] & -4.054 & 0.170 & 1363 & 1.001 \\
 $a$[3] & -3.694 & 0.177 & 1391 & 1.001 \\
 $b_{w}$[1] & 0.306 & 0.042 & 5546 & 0.999 \\
 $b_{w}$[2] & 0.376 & 0.027 & 4995 & 1.000 \\
 $b_{w}$[3] & 0.125 & 0.054 & 4790 & 1.000 \\
 $b_{l}$[1] & 0.511 & 0.129 & 4073 & 1.001 \\
 $b_{l}$[2] & 0.286 & 0.030 & 5803 & 1.000 \\
 $b_{l}$[3] & 0.126 & 0.026 & 4545 & 1.001 \\
 $\rho$[1] & 0.006 & 0.001 & 6015 & 1.000 \\
 $\rho$[2] & 0.004 & 0.000 & 7164 & 1.000 \\
 $\rho$[3] & 0.007 & 0.001 & 6386 & 1.000 \\
 $\sigma$ & 0.352 & 0.094 & 1967 & 1.001 \\
 $\delta$ & 0.839 & 0.294 & 1568 & 1.002 \\
 \hline\hline
\end{tabular}
\caption{Summary table for the main parameters of the $M_{full}$ model. The effective sample size (N. Eff.) and Rhat statistics very close to 1.0 are indications of a good sampling behaviour.}
\label{table:stats_model_full}
\end{table*}

\begin{figure*}[!h]
    \centering
    \includegraphics[width=1\textwidth]
    {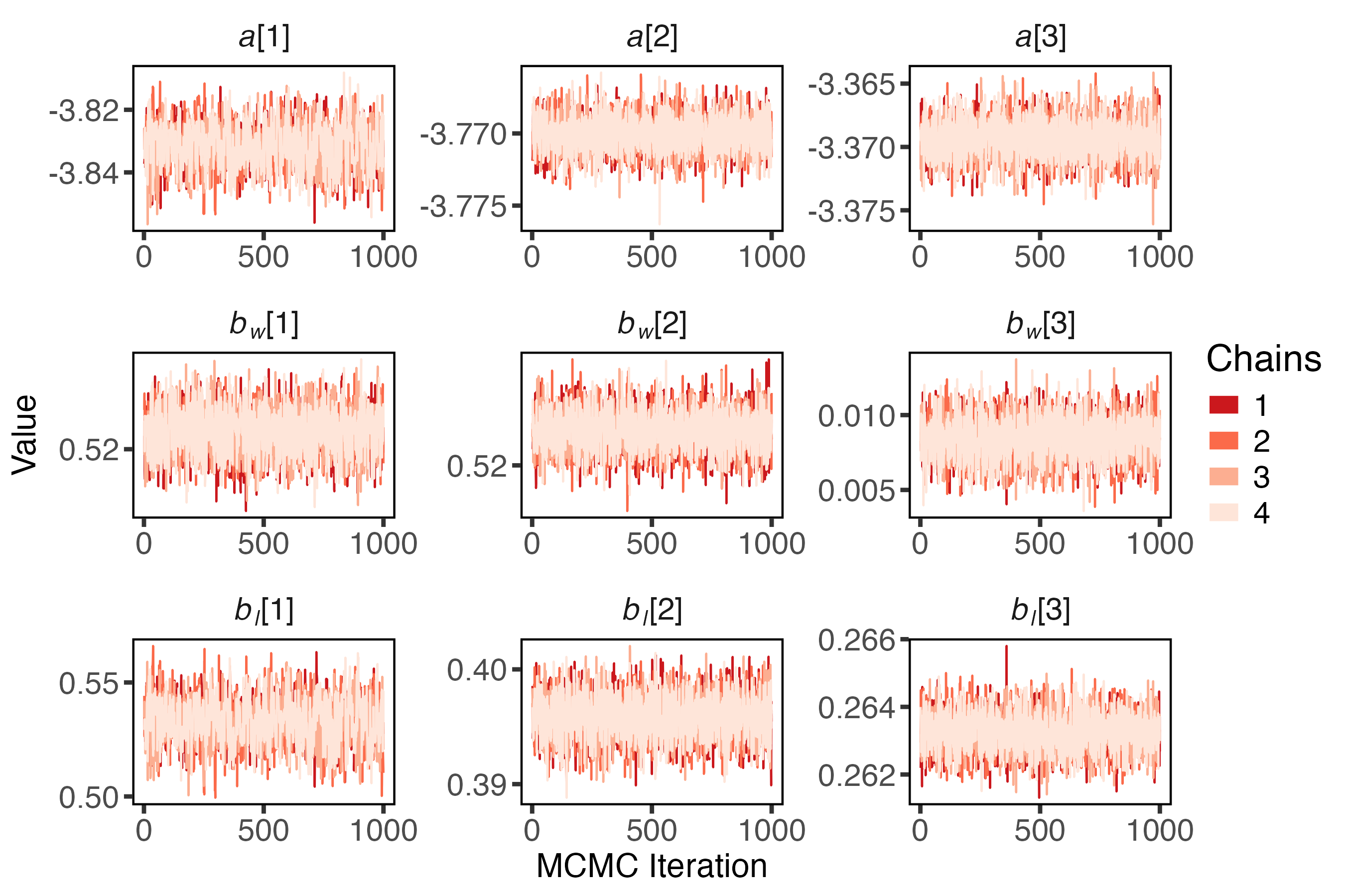}
    \caption{Trace plots of the sampled values for the main parameters of $M_{bin}$ at each iteration of the MCMC algorithm. All parameters show a healthy sampling behaviour.}
    \label{fig:traces_no_sp}
\end{figure*}

\begin{figure*}[!h]
    \centering
    \includegraphics[width=1\textwidth]
    {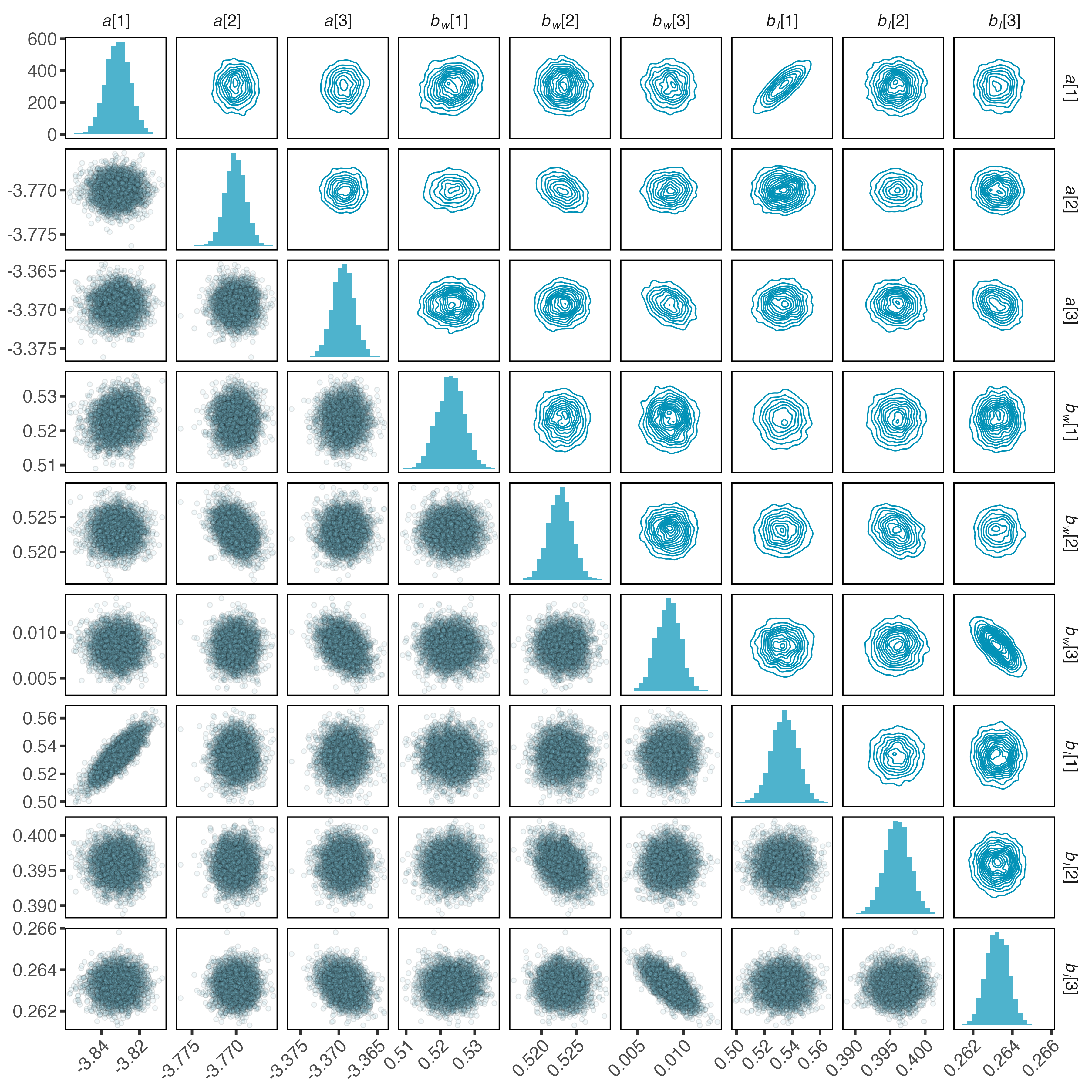}
    \caption{Pairwise scatter and contour plots of MCMC posterior samples obtained from $M_{bin}$. Moderate correlation is observed between the intercept and nighttime luminosity slope for rural areas and between working age population and nighttime luminosity slopes for urban areas.}
    \label{fig:pair_plot_no_sp}
\end{figure*}

\begin{figure*}[!h]
    \centering
    \includegraphics[width=1\textwidth]
    {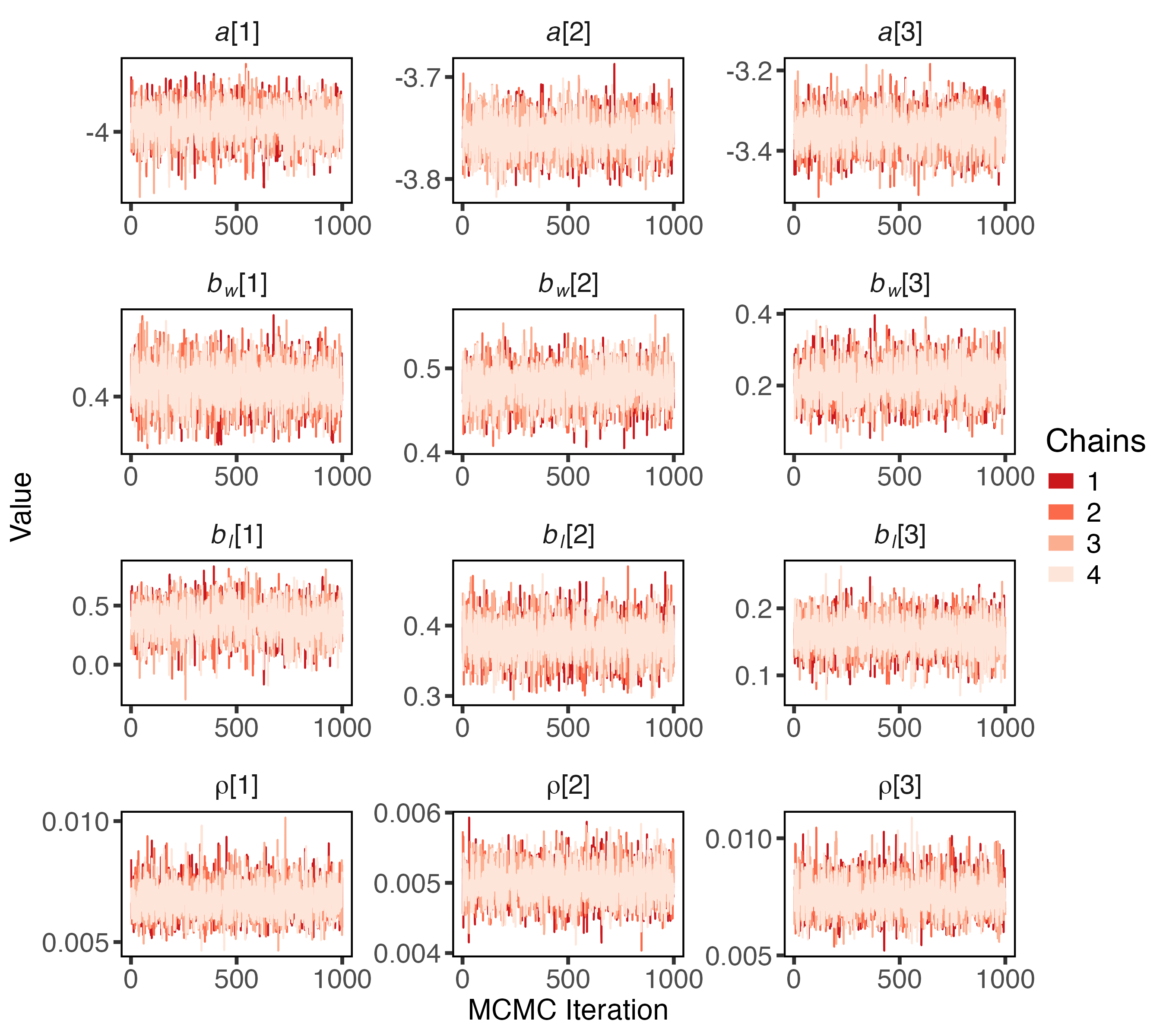}
    \caption{Trace plots of the sampled values for the main parameters of $M_{betabin}$ at each iteration of the MCMC algorithm. All parameters show a healthy sampling behaviour.}
    \label{fig:traces_no_sp_over}
\end{figure*}

\begin{figure*}[!h]
    \centering
    \includegraphics[width=1\textwidth]
    {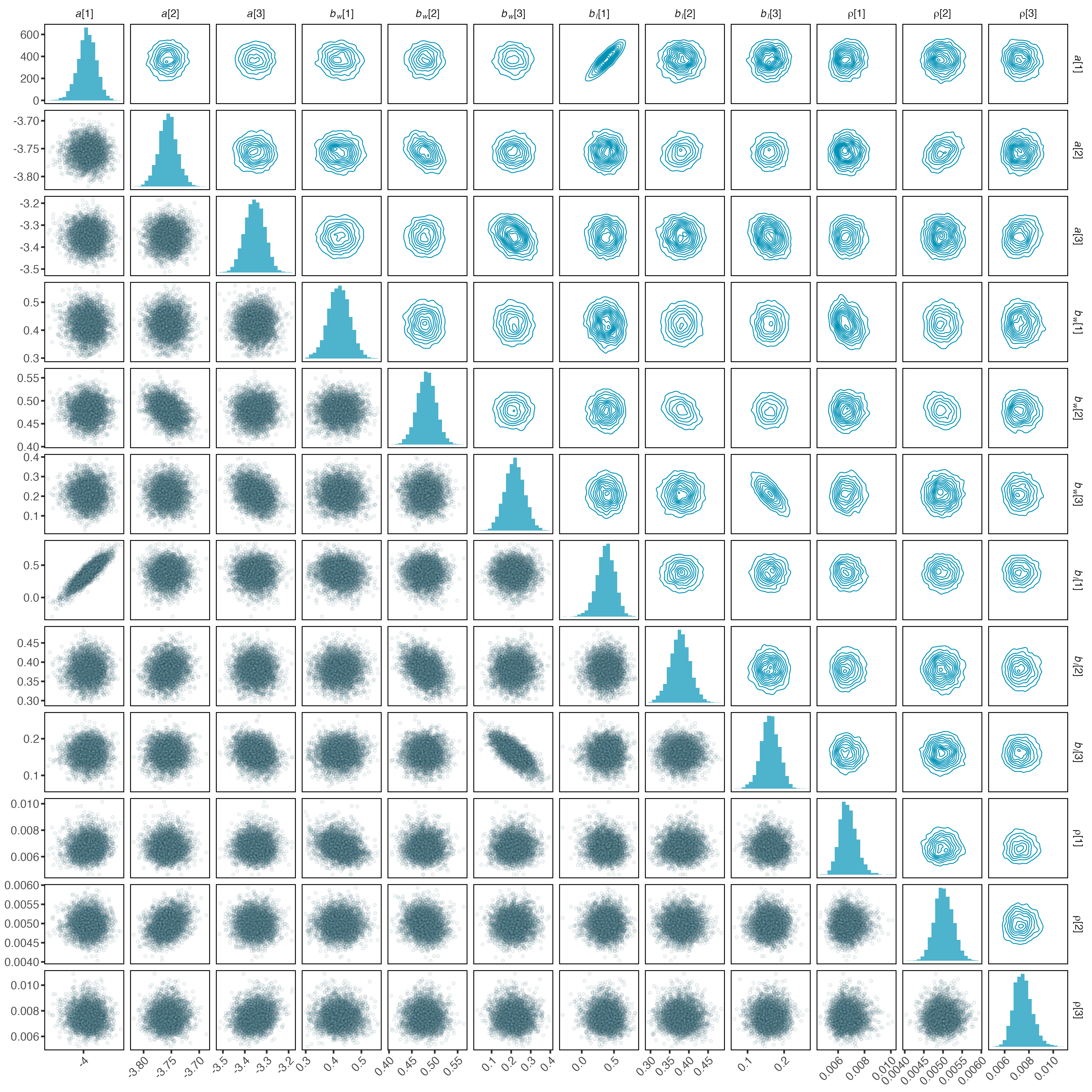}
    \caption{Pairwise scatter and contour plots of MCMC posterior samples obtained from $M_{betabin}$. Parameter correlations similar to those observed for $M_{bin}$ are also seen here.}
    \label{fig:pair_plot_no_sp_over}
\end{figure*}

\begin{figure*}[!h]
    \centering
    \includegraphics[width=1\textwidth]
    {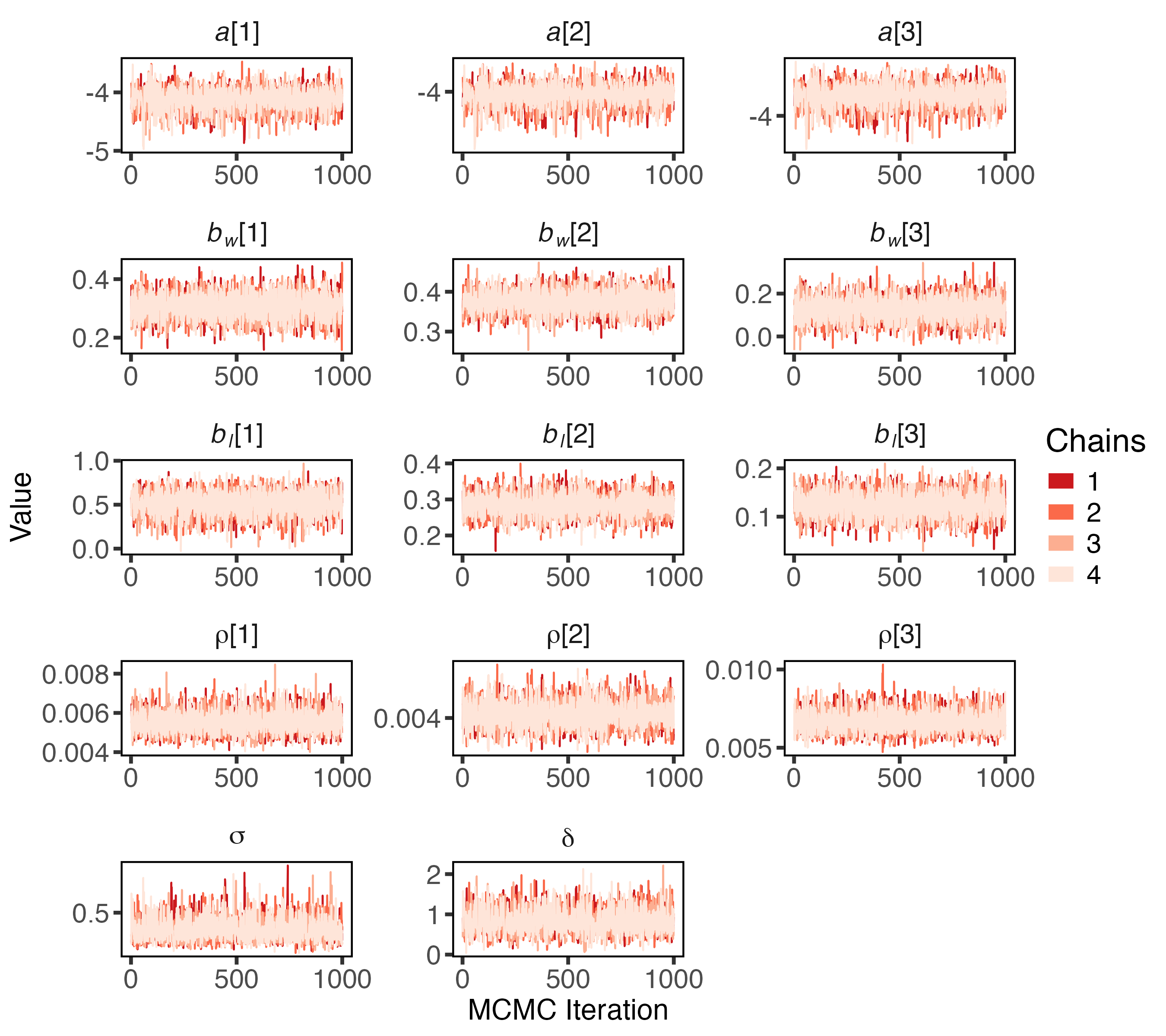}
    \caption{Trace plots of the sampled values for the main parameters of $M_{full}$ at each iteration of the MCMC algorithm. All parameters show a healthy sampling behaviour.}
    \label{fig:traces_full}
\end{figure*}

\begin{figure*}[!h]
    \centering
    \includegraphics[width=1\textwidth]
    {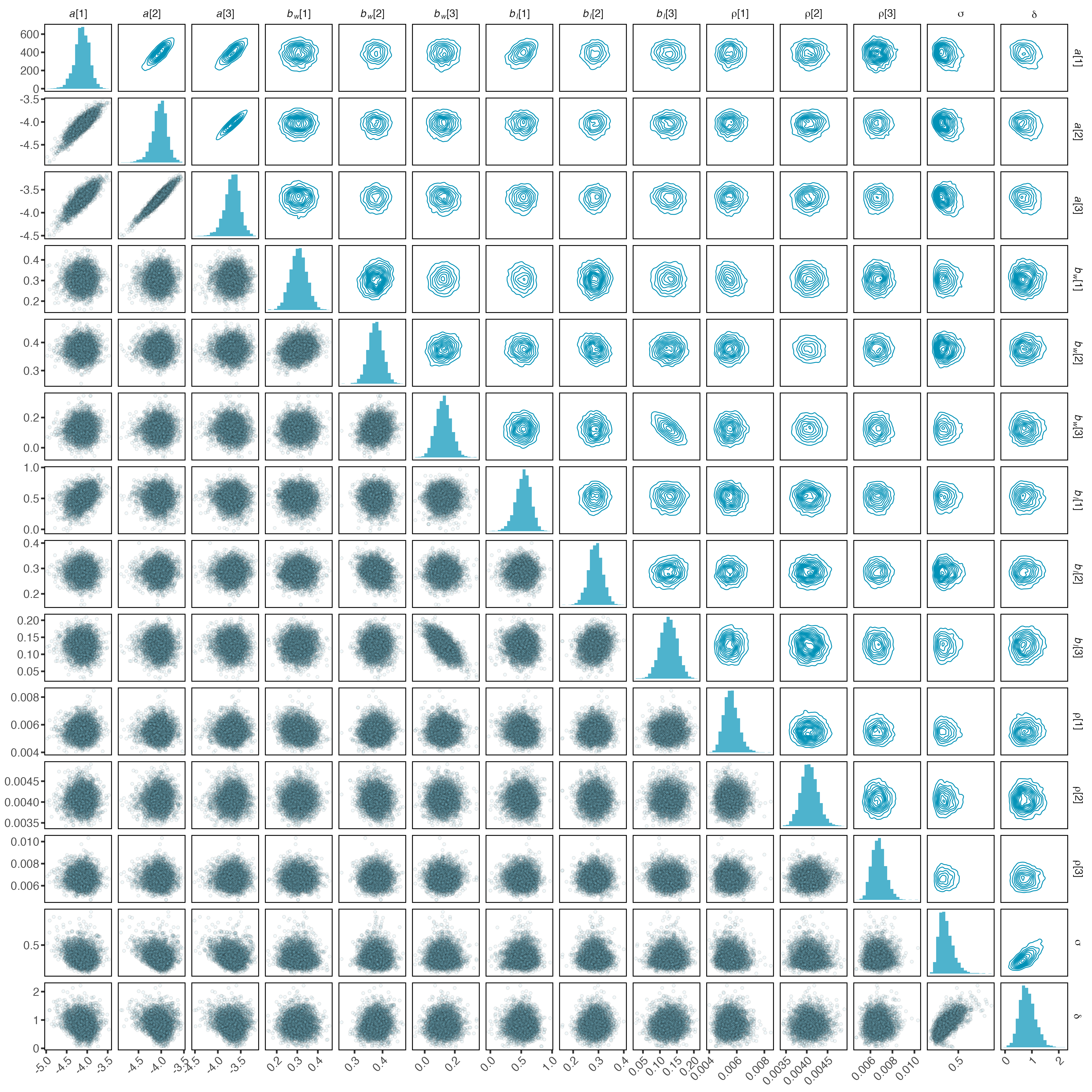}
    \caption{Pairwise scatter and contour plots of MCMC posterior samples obtained from $M_{full}$. The addition of an explicit spatial correlation term in the model induces strong correlations across the DUC dependent intercept terms. A sizable correlation is observed also between the two Gaussian Processes parameters ($\sigma$ and $\delta$).}
    \label{fig:pair_plot_full}
\end{figure*}

\begin{figure*}[!h]
    \centering
    \includegraphics[height=0.43\textheight]
    {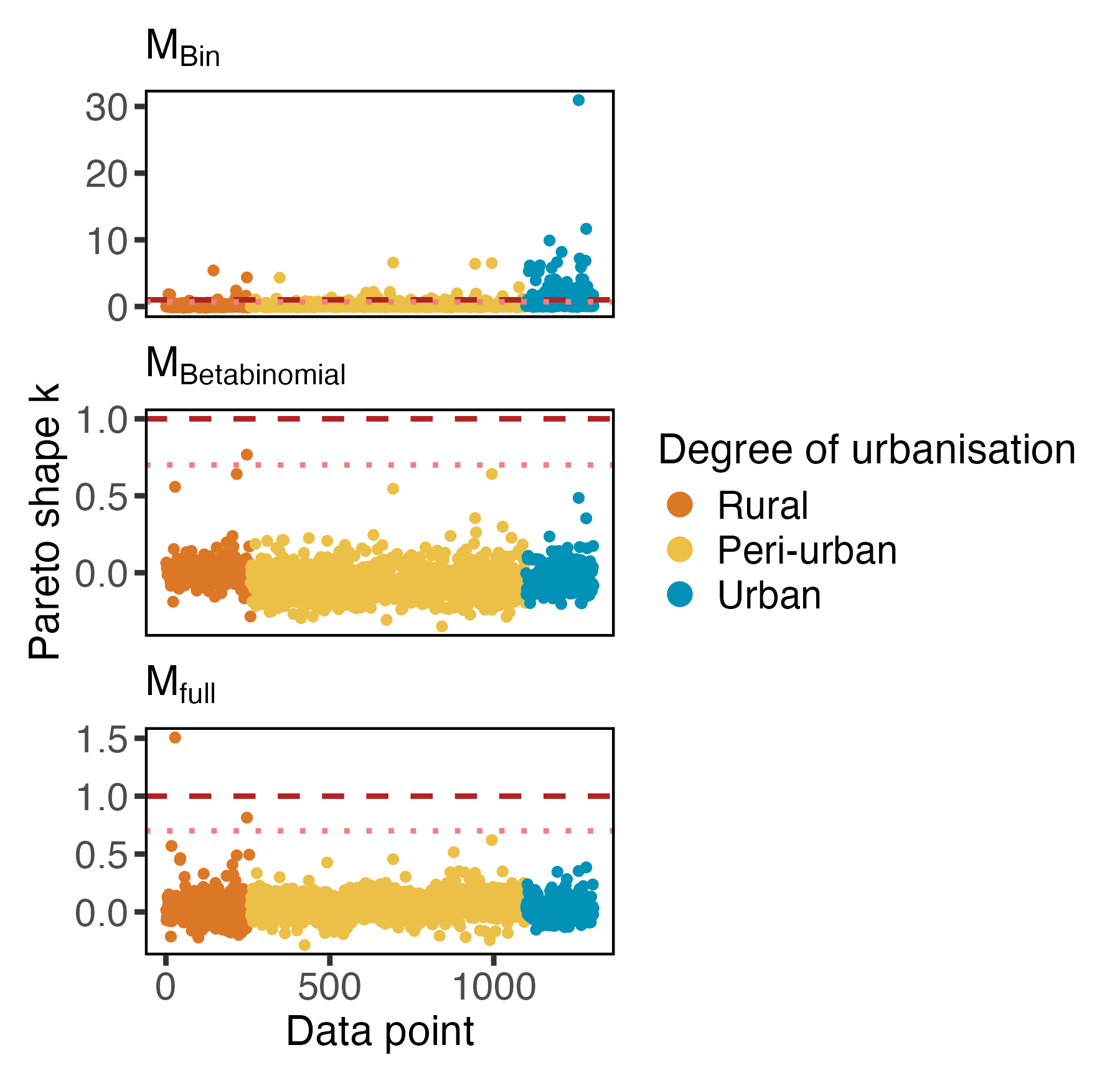}
    \caption{Pareto k diagnostics plots for the three Facebook uptake models. For $M_{bin}$ a large number of k values clearly exceed the 0.7 threshold, a hint that the model is not well-specified\cite{Vehtari2017}.}
    \label{fig:psis_diagnostics}
\end{figure*}

\begin{figure*}[!h]
    \centering
    \includegraphics[height=0.85\textheight]
    {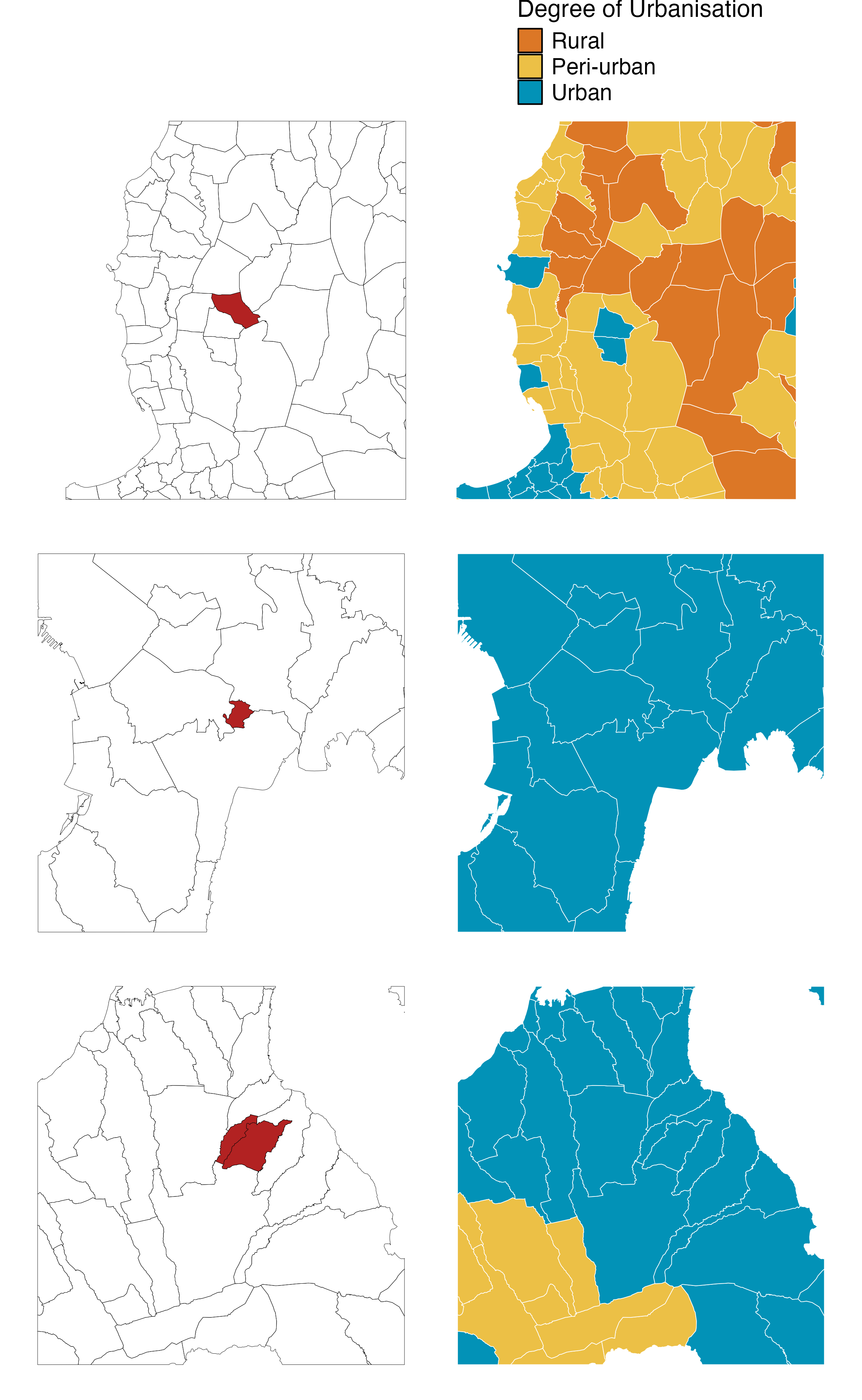}
    \caption{Detailed maps of the surroundings of areas with anomalous observed proportions of Facebook users (colored in red in the plots on the left). On the right the degree of urbanisation of all the administrative units in the region is shown. All the anomalous Bayans are small when compared to the size of the Bing tiles and are in close proximity to high population density urban areas.}
    \label{fig:outliers}
\end{figure*}

\clearpage

% \putbib[biblio]
% \end{bibunit}
\printbibliography
\end{refsection}

\end{document}